\documentclass{bioinfo}
\copyrightyear{}
\pubyear{2013}

\usepackage{fix-cm}
\usepackage{amsmath}
\usepackage{ctable}
\usepackage{tabularx}
\usepackage{multirow}
\usepackage{subfigure}
\usepackage{comment}
\usepackage{xr}
\usepackage{epsfig}

\begin{document}
\firstpage{1}

\title[Dynamic networks reveal key players in aging]{Dynamic networks reveal key players in aging}
\author[Faisal and Milenkovi\'{c}]{Faisal, F. E. and Milenkovi\'c, T.\footnote{To whom correspondence should be addressed}}
\address{Computer
  Science and Engineering, 
%ECK Institute for Global Health, and Interdisciplinary Center for Network
%  Science and Applications \\
  University of Notre Dame, Notre Dame, IN 46556, USA}

\history{}

\editor{}

\maketitle

\begin{abstract}
 
\section{Motivation:} Since susceptibility to diseases increases with age, 
studying aging gains importance.  Analyses of gene expression or
sequence data, which have been indispensable for investigating aging,
have been limited to studying genes and their protein products in
isolation, ignoring their connectivities.  However, proteins function
by interacting with other proteins, and this is exactly what
biological networks (BNs) model.  Thus, analyzing the proteins' BN
topologies could contribute to understanding of aging. Current methods
for analyzing systems-level BNs deal with their static
representations, even though cells are dynamic.  For this reason, and
because different data types can give complementary biological
insights, we integrate current static BNs with aging-related gene
expression data to construct dynamic, age-specific BNs. Then, we apply
sensitive measures of topology to the dynamic BNs to study cellular
changes with age.

\section{Results:} While global BN topologies do not significantly change 
with age, local topologies of a number of genes do. We predict such
genes as aging-related. We demonstrate credibility of our predictions
by: 1) observing significant overlap between our predicted
aging-related genes and ``ground truth'' aging-related genes; 2)
showing that our aging-related predictions group by functions and
diseases that are different than functions and diseases of genes that
are not predicted as aging-related; 3) observing significant overlap
between functions and diseases that are enriched in our aging-related
predictions and those that are enriched in ``ground truth''
aging-related data; 4) providing evidence that diseases which are
enriched in our aging-related predictions are linked to human aging;
and 5) validating all of our high-scoring novel predictions via manual
literature search.

%10\% of our predictions in the ``ground truth'' data or in the literature.

\section{Contact:} \href{tmilenko@nd.edu}{tmilenko@nd.edu}

%\section{Supplementary information:}
%Available at \emph{Bioinformatics} online.

\end{abstract}

\vspace{-0.5cm}

\section{Introduction}
\label{sect:intro}

\subsection{Motivation and background}
\label{sect:motivation}
 
Since the US is on average growing older because of $\sim$78 million
of baby boomers who have began turning 65 in 2011, and since
susceptibility to diseases increases with age, studying human aging
gains importance.  Analysis of gene expression data has been
indispensable for investigating aging
\citep{Wieser2011,Fortney2010}. However, it has mostly been limited to
studying differential expression of individual genes, without
considering their connectivities \citep{Fortney2010}. But, it is the
proteins (gene products) that carry out cellular processes and they do
so by interacting with other proteins instead of acting alone.  And
this is exactly what biological networks and \emph{protein-protein
interaction (PPI) networks} in particular model; in PPI networks,
nodes are proteins and edges correspond to physical interactions
between the proteins.  Thus, analyzing topologies of proteins in PPI
networks could contribute to understanding of the processes of aging.
Although as a proof of concept this study focuses on PPI networks, it
is applicable to other types of biological networks.  High-throughput
screens for PPI detection have yielded systems-level (though
incomplete) PPI networks for many organisms,
%\citep{GiotSci03,Stelzl05,Yu2008,Simonis2009}, 
which are publicly available \citep{BIOGRID}.

The majority of current methods for analyzing systems-level PPI
networks deal with their \emph{static} representations, due to
limitations of biotechnologies for PPI data collection, even though
cells are dynamic \citep{Przytycka2010}. For this reason, and because
different data types can give complementary biological insights
\citep{Memisevic10b,Przytycka2010}, we \emph{integrate} current static
PPI network data \citep{HPRD,BIOGRID} with age-specific gene
expression data \citep{Lu2004} to computationally construct
\emph{dynamic, age-specific PPI networks}, in order to study cellular
changes with age from such networks.

Furthermore, topological positions of aging-related genes in the
\emph{static} networks have been studied
\citep{Kriete2011,aging1,Reja2009,aging2,Magalhaes2009}, but mostly
with \emph{crude} measures of topology that can not cope with the
complexity of PPI networks \citep{Przulj2011}. For example, node
degrees have been used to argue the central role or aging-related
proteins in the yeast network compared to proteins that are not
associated with aging, or to study the role of chaperones (heat stock
proteins) in aging \citep{aging2,Soti2007}.  In addition to aging,
many approaches have aimed to link node degrees with, for example,
essentiality \citep{Jeong01}, disease
\citep{Sharan2008,Sharan10}, cancer \citep{JB06,Aragues2008},
or pathogenicity \citep{Dyer2008}.  However, it is possible that the
high-degree proteins have been more studied simply because of their
known relevance to human health \citep{Przulj2011,Ratman2009}. Hence,
more constraining measures of topology might be needed that go beyond
capturing only the direct network neighborhood of a node
\citep{Milenkovic2008,MMGP_Roy_Soc_09}.  While such measures
exist and have been used to link proteins' network positions with
their involvement in some biological processes
\citep{Sharan2007,Milenkovic2011}, to our knowledge, they have
not been linked to proteins' involvement in aging even in static and
especially in dynamic PPI networks.  Here, we apply \emph{a series} of
measures of topology, including some highly sensitive measures
\citep{Milenkovic2008,Milenkovic2011}, to the
\emph{dynamic} PPI networks to identify key players in aging.

\begin{figure*}
  \begin{center} \begin{minipage}[t]{0.72\linewidth}
\vspace{-1.1cm}
  \raisebox{-6.6cm}{\epsfig{file=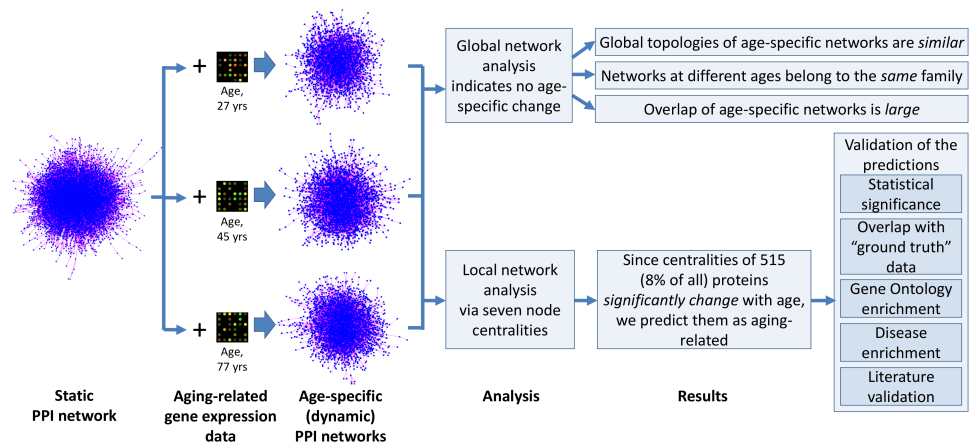,
  width=\linewidth}} \end{minipage}\hfill
\begin{minipage}[t]{0.235\linewidth} \caption{Summary of our study. We integrate a static PPI network with
  aging-related gene expression data to obtain age-specific networks.
  We analyze changes in global and local network topologies with age.
  While global network analysis indicates no age-specific change,
  local topologies (as captured by seven node centrality measures) of
  some proteins do significantly change with age. We predict such
  proteins as aging-related and validate our predictions in several
  ways. 
  \label{fig:front-diagram} } \end{minipage} \end{center}
  \vspace{-0.4cm}
\end{figure*}

%\begin{figure*}[tb]
%\centering
%\includegraphics[scale=0.13]{figures/fig-front-diagram}
%\vspace{-0.45cm}
%\caption{Summary of our study. We integrate a static PPI network with
%  aging-related gene expression data to obtain age-specific networks.
%  We analyze changes in global and local network topologies with age.
%  While global network analysis indicates no age-specific change,
%  local topologies (as captured by seven node centrality measures) of
%  some proteins do significantly change with age. We predict such
%  proteins as aging-related and validate our predictions in several
%  ways. }
%\label{fig:front-diagram}
%\end{figure*}

\vspace{-0.7cm}

\subsection{Our study}
\label{sect:ourstudy}

We aim to study human aging via integration of aging-related gene
expression data with static PPI network data (Fig.
\ref{fig:front-diagram}). We obtain dynamic, age-specific PPI networks
by selecting in the static network: 1) all proteins that correspond to
actively expressed genes at different ages and 2) all PPIs involving
these ``active'' proteins.  Hence, each age-specific
%PPI 
network is the network that is ``active'' at a given age.  We
hypothesize that the dynamic and integrative network analysis provides
a valuable model of cellular functioning that can reveal aging-related
information and that can reveal more of the information than static
analysis of individual data types.

Given the dynamic network data, we first aim to answer whether the
overall network topologies change with age (Fig.
\ref{fig:front-diagram}). 
%We do so by comparing
%\emph{global} network properties of the  age-specific networks, 
%by evaluating the fit of each of the age-specific networks to a series
%of well-known graph families \citep{GraphCrunch}, and by measuring the
%overlap between the age-specific networks. We find that the global
%network topologies do \emph{not} significantly change with age.
%global properties of the networks do not vary with age, that networks
%at different ages belong to the same graph family, and that the
%overlap between the networks is quite large.
%
Since this is \emph{not} the case,
%Given this,
%Since the global network topologies do not change with age, and since
and since the gene expression data alone revealed only a small portion
of all genes as aging-related \citep{Lu2004}, it could be that
\emph{local} topologies around only a \emph{subset} of proteins in
the network do change with age. Hence, we study positions of proteins
in each age-specific PPI network with respect to measures of local
topology, called \emph{node centralities}, with the goal of
identifying proteins whose centralities significantly change with age.
% with respect to at least one measure.  
We find 515 such proteins (8.1\% of all proteins in the static 
network), which is quantitatively consistent to the result by
\cite{Lu2004}. We predict these proteins as aging-related and 
%We study relationships between the different node
%centrality measures. In this context, we score our predictions so that
%the more centrality measures support a prediction and the higher the
%significance of the change of centrality values with age, the more
%credible the prediction. 
%
%We 
validate them as follows.

%\vspace{-0.35cm}
%
%\begin{itemize}
%%\setlength{\itemindent}{-1em}
%
%\item 
\hspace{-0.3cm}\textbf{1)} The predictions are statistically significant, i.e., non-random.
%
%\vspace{-0.1cm}

%\item 
\hspace{-0.3cm}\textbf{2)} The overlap of our predicted aging-related genes and ``ground
truth'' aging-related genes is significant. All of our five
\emph{highest-scoring} predictions, namely GORASP2, MAP2K4, TIAM1,
MAP1B, and S100B, are present in \emph{multiple} aging-related
``ground truth'' data sets.  Nonetheless, many of our predictions are
\emph{novel}, i.e., absent from the ``ground truth'' data.  This 
confirms  that dynamic network analysis of integrated
data types can reveal additional biological knowledge compared to
static analysis of individual data.

%\vspace{-0.1cm}

%\item 
\hspace{-0.3cm}\textbf{3)} Our predictions  group by biological functions 
and diseases that are different than functions and diseases of genes
that we do not predict as aging-related. The overlap between functions
and diseases that are enriched in our predictions and those that are
enriched in the ``ground truth'' aging-related data is significant.
Diseases that are enriched in our predictions are linked to human
aging.

%\vspace{-0.1cm}

%\item 
\hspace{-0.3cm}\textbf{4)} We \emph{manually} search in the literature for our top 10\%
highest-scoring predictions that are not present in the ``ground
truth'' aging-related data, and we successfully validate all of them.

%\end{itemize}

\vspace{-0.4cm}

\begin{methods}
\section{Materials and methods}

\subsection{Data}
\label{sect:data}

\subsubsection{Aging-related gene expression data}
\label{sect:humanagingbrain}

We use human brain gene expression data consisting of 30 samples
obtained from 30 individuals between 26 and 106 years of age.
%, with one sample per individual \citep{Lu2004}.  
In an individual, \cite{Lu2004} defined a gene as being
\emph{expressed} (or active) at a given age if its \emph{detection
$p$-value}, which indicates the significance level of its mRNA
abundance at that age, is less than $0.04$.  We adopt the same
procedure (Supplementary Section
\ref{supple:supplesect:humanagingbrain}).

\subsubsection{Static PPI network data}
\label{sect:ppinetworks}

We obtain human static PPIs from HPRD \citep{HPRD} and BioGRID
\citep{BIOGRID}.  HPRD data consists of 9,322 unique proteins and
36,030 unique PPIs between the proteins.  BioGRID data consists of
10,078 unique proteins (with respect to their gene IDs) and 50,954
unique PPIs between the proteins.

\subsubsection{Integrating static PPI network with gene expression
  data to form age-specific PPI networks}
\label{sect:dynamicppinetworks}

We form dynamic, age-specific networks as follows.  To form the
network specific to a given age, we select in the static network those
proteins that are expressed at that age (Section
\ref{sect:humanagingbrain}) and all PPIs that exist between the
expressed proteins (see Supplementary
Section~\ref{supple:supplesect:dynamicppinetworks} for a formal
description). Since gene expression data is collected for 30 ages, 30
age-specific networks can be formed from the given static network.
Since we study two static networks (HPRD and BioGRID), we obtain two
sets of dynamic networks. We run subsequent analyses on each of the
network sets.  Since we find that results are similar across the two
sets, for simplicity, here we report results only for the HPRD
network.  Results for the BioGRID network are reported in the
Supplement.

\subsubsection{``Ground truth'' aging-related
  data}\label{sect:ground_truth}

We denote the set of genes present in both the static PPI network
(Section \ref{sect:ppinetworks}) and brain gene expression data
(Section \ref{sect:humanagingbrain}) as \emph{StatNetExpression}.

By studying brain gene expression data from Section
\ref{sect:humanagingbrain}, \cite{Lu2004} predicted 442 genes as
aging-related, as their expression significantly correlated with
age. Of these, 341 genes are present in StatNetExpression. Henceforth,
we denote this ``ground truth'' aging-related set of 341 genes
predicted from \emph{brain gene expression data alone} as
\emph{BrainExpression2004Age}.

By studying another brain gene expression data with
% consisting of
174 samples 
%taken 
from 55 individuals, with multiple sample per individual,
\cite{Berchtold2008} identified 8,277 genes (via 12,514 probes)
whose expression significantly changed with age. Of these, 3,228 are
present in StatNetExpression. Henceforth, we denote this ``ground
truth'' aging-related set of 3,228 genes predicted from
\emph{brain gene expression data alone} as \emph{BrainExpression2008Age}. 

Clearly, BrainExpression2004Age and BrainExpression2008Age are very
similar in the sense that their aging-related genes have been inferred
from brain gene expression data.  (And as such, among all ``ground
truth'' data sets (see below), these two sets are expected to be the
most similar to our aging-related predictions, since our predictions
are also partly based on brain human gene expression data.) However,
it is important to note that BrainExpression2004Age and
BrainExpression2008Age were predicted from two independent data sets,
and compared to BrainExpression2004Age, BrainExpression2008Age is a
result of a newer microarray study, it covers more samples and more
individuals, and it covers more samples per individual.

By studying brain gene expression data set related to different stages
of Alzheimer's disease (AD), 
%such as entorhinal, isocortical, or limbic stage, 
\cite{Simpson2011} identified 2,911 genes (linked to
3,404 probes) that have significantly different expression levels at
different stages of AD. Of these, 1,117 are present in
StatNetExpression. Henceforth, we denote this ``ground truth''
AD-related set of 1,117 genes predicted from \emph{brain gene
expression data} as \emph{ADExpressionAge}.

By studying gene expression data related to Hutchinson-Gilfold
progeria syndrome (HGPS), a human premature aging-related disease,
\cite{Liu2011} identified 1,731 genes that have a differentially
methylated region between wild-type and HGPS-affected fibroblasts of
vascular muscles cells. Of these, 708 are present in
StatNetExpression. Henceforth, we denote this ``ground truth''
HGPS-related set of 708 genes predicted from
\emph{vascular muscles-related gene expression data} as
\emph{HGPSExpressionAge}.

In July 2012, GenAge contained 261 human genes that have been linked
to aging as sequence-based orthologs of aging-related genes in model
species \citep{Magalhaes2009a}. Of these, 230 are present in
StatNetExpression.  Henceforth, we denote this ``ground truth''
aging-related set of 230 genes predicted from \emph{sequence data} by
\emph{SequenceAge}.

%Clearly, of all the five ground truth data sets,
%BrainExpression2004Age and BrainExpression2008Age are the most likely
%to be similar to our aging-related predictions (as all three are based
%on \emph{brain-related} data, brain-related \emph{gene expression}
%data, and brain \emph{aging-related} gene expression data), followed
%by ADExpressionAge (as our predictions as well as this data are both
%based on \emph{brain-related} data and brain-related \emph{gene
%expression} data), followed by HGPSExpressionAge (as our predictions
%as well as this data are both based on \emph{gene expression} data),
%and finally followed by SequenceAge (as our predictions and this data
%do not have any obvious connections).

%Note that we intentionally use quotes when talking about aging-related
%``ground truth'' data, since the data has been predicted
%computationally and \emph{not} obtained experimentally. This is
%because human aging is hard to study experimentally due to long life
%span and ethical constraints.

\subsubsection{Complements of the ``ground truth'' aging-related
  data}\label{sect:complements_ground_truth}

We define a set of genes as the \emph{complement} of a ``ground
truth'' aging-related data set if the genes are present in
StatNetExpression but not in the ``ground truth'' data set. We denote
the complements of BrainExpression2004Age, BrainExpression2008Age,
ADExpressionAge, HGPSExpressionAge, and SequenceAge as
\emph{BrainExpression2004Complement},
\emph{BrainExpression2008Complement}, \emph{ADExpressionComplement},
\emph{HGPSExpressionComplement}, and \emph{SequenceComplement},
respectively.

All above data sets are defined with respect to HPRD PPI data. For
BioGRID data, see Supplementary Section
\ref{supple:supplesect:ground_truth}.

\subsection{Do global network topologies change with age?}
\label{sect:methods_global_analysis}

Given the dynamic, age-specific PPI networks, we test whether the
overall (global) topologies of the networks change with age.  We do so
by comparing the different networks with respect to several commonly
used global network properties (Section
\ref{sect:methods_compare_global}), by evaluating the fit of each of
the age-specific networks to a series of well-known graph families,
i.e., network models (Section~\ref{sect:methods_models})
\citep{GraphCrunch, GraphCrunch2}, and by measuring the overlap of the
age-specific networks (Section \ref{sect:methods_overlap_network}).

\subsubsection{Comparing global properties of age-specific networks}\label{sect:methods_compare_global}

We analyze three properties: the average clustering coefficient,
average diameter, and graphlet frequency distribution
\citep{Memisevic10a}.  The properties are
% former two are 
defined in Supplementary Section
\ref{supple:supplesect:methods_compare_global}.
%The \emph{graphlet frequency distribution} of a network
%counts the frequency of occurrence of 3-5-node graphlets in the
%network; a \emph{graphlet} is an induced subgraph \citep{Przulj04}. 

\subsubsection{Evaluating the fit of age-specific networks to
  different graph families or network models}
\label{sect:methods_models}

We compare the fit of the dynamic PPI networks to different graph
families, i.e., network models \citep{Milenkovic2009}, to test whether
the best fitting model changes with age.  Various network models have
been proposed. We use: (1) Erd\"{o}s-R\'{e}nyi random graphs (ER), (2)
generalized Erd\"{o}s-R\'{e}nyi random graphs with same degree
distribution as the data (ERDD), (3) geometric random graphs (GEO),
(4) geometric gene duplication and mutation model (GEOGD), (5)
scale-free networks (SF), and (6) scale-free gene duplication and
mutation model (SFGD) \citep{GraphCrunch,GraphCrunch2}.  To evaluate
the fit of the data network to a given model, we compare the topology
of the data network to the topology of a random network instance drawn
from the model with respect to a highly constraining measure of
network topological similarity called \emph{graphlet degree
distribution agreement (GDD-agreement)}
\citep{Przulj06ECCB}. For details, see Supplementary Section \ref{supple:supplesect:methods_models}.

\subsubsection{Computing the overlap between age-specific networks}
\label{sect:methods_overlap_network}

We measure the overlap between each pair of age-specific networks as
the percentage of nodes (or edges) in the smaller of the two networks
that are common to the two networks.  For details, see Supplementary
Section \ref{supple:supplesect:methods_overlap_network}.

\subsection{Do local topologies of proteins change with age?}
\label{sect:methods_local_analysis}

We study topological positions of proteins in each age-specific
network with respect to seven node centrality measures (Section
\ref{sect:methods_NC}). We predict as aging-related those proteins
whose centralities significantly change with age (Section
\ref{sect:corr-age-cent}).  We validate our predictions in several
ways (Section \ref{sect:methods_validation}).

\subsubsection{Local measures of topology or node
  centralities}\label{sect:methods_NC}

%Different node centralities capture different aspects of the network
%position of a node.  We use seven different centrality measures, as
%follows.

Various centrality measures have been used to link topological
importance of a node in the network to its functional importance.
Below, we define each of the seven measures that we use and provide
biological justification for their use.

\emph{Degree centrality} (DEGC) measures the degree of a node in the
network, i.e., the number of the node's neighbors. 
%Intuitively, 
The higher the degree of a node, the more central the node according
to DEGC.  Since current PPI networks have ``power-law'' degree
distributions, with many low-degree nodes and few high-degree nodes,
and since removal of the high-degree nodes would impact the network
structure (by disconnecting it), DEGC of a gene has been related to the
gene's essentiality as well as its involvement in disease
\citep{Barabasi_Oltvai04,Sharan2008}.

\emph{Clustering coefficient centrality} (CLUSC) measures, for a given
node, how many pairs of neighbors of the node are connected by an
edge, out of all pairs of the node's neighbors. Intuitively, the more
interconnected the neighborhood of the node, the more central the node
is according to CLUSC.  In a PPI network, a node with high clustering
coefficient, together with the node's neighbors, forms a highly
interconnected network region, which is likely to correspond to a
functional module \citep{Barabasi_Oltvai04}.

\emph{$K$-core} of a network is a maximal subset of nodes in the
network such that each node is connected to at least $k$ others in the
subset. \emph{$K$-coreness centrality} (KC) of a node is $k$ if the
node is in $k$-core.  Nodes with high KC in the human PPI network have
been found to correspond to ``core diseaseome,'' a subnetwork that is
significantly enriched in disease genes and drug targets
\citep{Janjic2012}, as well as to influential ``spreaders'' of
information throughout the network \citep{Kitsak2010}.

\emph{Graphlet degree centrality} (GDC) measures how many graphlets a
node participates in, for all 2-5-node graphlets
\citep{Milenkovic2011}. Intuitively, the more graphlets a node
touches, the more central the node is according to GDC. Since it
captures the \emph{extended} network neighborhood of a node,
% and since
%it does so by relying on the mathematically constraining notion of
%graphlets, 
GDC is a highly sensitive measure of network topology.  Thus, in a PPI
network, proteins with high GDCs represent potential candidates for
therapeutic intervention, since targeting such proteins with drugs
would have more significant impact on the network structure than
targeting proteins that reside in sparse and non-complex network
regions \citep{Milenkovic2011}. Indeed, GDC has been found to capture
well disease and pathogen-interacting proteins and drug targets
\citep{Milenkovic2011}.

\emph{Betweenness centrality} (BETWC) measures the involvement of a
node in the shortest paths in the network. Intuitively, nodes that
occur in many shortest paths have high centrality according to
BETWC. BETWC of node $v$, $C_\textrm{betwc}(v)$, is:
$C_\textrm{betwc}(v)={\displaystyle {\displaystyle \sum_{s\neq v\neq
t\in V}{\displaystyle
\frac{\sigma_{st}(v)}{\sigma_{st}}}}}$, where $V$ is the set of nodes
in the network, $\sigma_{st}$ is the number of shortest paths between
nodes $s$ and $t$, and $\sigma_{st}(v)$ is the number of shortest
paths between $s$ and $t$ that go through $v$.  In a PPI network,
BETWC of a protein indicates the ``likelihood'' of the protein to
participate in pathways connecting all other proteins
\citep{Koschutzki2008}. Removal of a protein that is on
critical pathways between many other proteins could cause loss of
communication between the proteins.  Also, targeting such a node with
a drug could cause the drug effects to spread fast to all the nodes
\citep{Milenkovic2011}.  This property has been used to identify
gene-disease associations by encoding each gene in the network based
on the distribution of shortest path lengths to all genes associated
with disease \citep{Radivojac2008}. Also, see
%BETWC has been used in other contexts as well
\cite{Kitsak2010}.

\emph{Closeness centrality} (CLOSEC) measures the ``closeness'' of a
node to all other nodes in the network.  Intuitively, nodes with small
shortest path distances to all other nodes have high centrality
according to CLOSEC. CLOSEC of node $v$, $C_\textrm{closec}(v)$, is:
$C_\textrm{closec}(v)=\frac{1}{\sum_{u\in{V}}\sigma(u,v)}$, where
$\sigma(u,v)$ is the shortest path distance between nodes $u$ and $v$.
%It is sometimes
%normalized by the number of nodes in the network, $|V|$, as:
%$C_c(v)=\frac{|V|}{\sum_{u\in{V}}d(u,v)}$.  
In a PPI network, CLOSEC of a protein indicates the ``likelihood'' of
the protein to reach or be reachable from all other proteins
\citep{Scardoni2009}. And it has been a widely accepted assumption that
proteins that are closer to each other are more likely to perform the
same function \citep{Sharan2007}.

\emph{Eccentricity centrality} (ECC) is very related to CLOSEC, except
that it measures the ``closeness'' of a node \emph{only} to the
\emph{farthest} node in the network
\citep{Wuchty2003}. Intuitively, nodes with small
shortest path distances to the furthest node in the network have high
centrality according to ECC. ECC of node $v$, $C_\textrm{ecc}(v)$, is:
$C_\textrm{ecc}(v)=\frac{1}{\max_{u\in{V}}\{\sigma(u,v)\}}$.
%In a PPI network,
%ECC of a protein indicates the ``likelihood'' of the protein to reach
%or be reachable from the farthest protein

\subsubsection{Prediction of aging-related genes}
\label{sect:corr-age-cent}

For each measure, we compute centrality values for a node in each of
the 30 age-specific networks. Then, we calculate Pearson or Spearman
correlation between the 30 ages and the node's 30 centrality values
(Supplementary Section \ref{supple:supplesect:corr-age-cent}).  We do
this for all genes that are expressed in at least five  ages
in
\cite{Lu2004}'s brain gene expression data (Section
\ref{sect:humanagingbrain}).  If such a gene is unexpressed at a given
age, we assign it a centrality value of zero at that age.  Since
results are consistent for both correlation measures, here we report
results only for Pearson correlation.  Results for Spearman
correlation are shown in the Supplement.

We quantify the statistical significance of the given correlation
value by measuring the probability (i.e., $p$-value) of observing by
chance a better value (i.e., the same or higher value when the
original value is positive, or the same or lower value when the
original value is negative). We do this by randomly reshuffling the 30
node centrality values at the 30 ages and by computing the resulting
``random correlation''.  We repeat this 1,000 times to get 1,000
random correlations. We compute the $p$-value as the percentage of the
1,000 runs in which the random correlation is better than the original
one.  We predict a gene as aging-related if its $p$-value is below
0.01.
%%%
%We argue that multiple test correction is not applicable in
%determining $p$-value for each gene because we do not experimentally
%determine whether a gene is aging-related. Rather we use computational
%method to predict aging-related genes and assign prediction scores
%indicating the credibility of them to be aging-related.
%%%

Since we study multiple node centralities, each of which can predict
the given gene as aging-related, we score our predictions so that the
more centrality measures support a prediction and the higher the
significance of the change of its centrality values with age, the
higher the score and the more credible the prediction. For details,
see Supplementary Section
\ref{supple:supplesect:corr-age-cent}.

\subsubsection{Validation of predicted aging-related genes}
\label{sect:methods_validation}

\emph{ }

\vspace{0.1cm} \hspace{-0.33cm}\textbf{Statistical significance of our
predictions.} 
%Before we predict a gene as aging-related, we ensure
%that the correlation between its centrality values and age is
%statistically significant (Section \ref{sect:corr-age-cent}). Also, 
To test whether our approach of combining static network data with
aging-related expression data into the dynamic network data actually
gives meaningful predictions, we study whether the number of
aging-related genes that we predict from the actual data is
statistically significantly larger than the number of aging-related
genes that we predict from ``randomized data''. By ``randomized
data'', we mean that we randomize the expression data before
integrating it with the static network data (Supplementary Section
\ref{supple:supplesect:method-randomize}).  Then,  we integrate the
randomized expression data with the static PPI network, construct 
randomized age-specific networks just as in Section
\ref{sect:dynamicppinetworks}, and predict aging-related genes from 
the randomized networks just as in Section
\ref{sect:corr-age-cent}. We repeat the above procedure multiple
times, in order to assign a $p$-value to the number of predictions
that we make from the actual data (Supplementary Section
\ref{supple:supplesect:method-randomize}).

\vspace{0.1cm} \hspace{-0.33cm}\textbf{Overlap between genes of different data sets.} We measure the statistical significance of the overlap of
  genes in one data set 
%(e.g., our aging-related predictions) 
and genes in another data set 
%(e.g., a ``ground truth'' data set) 
by using the hypergeometric test, 
%model for sampling without
%replacement. The test 
which computes probability $p$ (i.e., $p$-value) of observing the same
or larger overlap by chance as follows. Let $E$ denote the set of
genes that are present in StatNetExpression. $|E|=6,397$.  Let $A$
denote the subset of genes in $E$ that are in any one of the two data
sets.  Let $G$ denote the subset of genes in $E$ that are in the other
data set. Let $O$ denote the set of genes that are in the overlap
between $A$ and $G$.  Then, $p$ is:
$p=1-\displaystyle\sum_{i=0}^{|O|-1} \dfrac{{|E| \choose i}{|E|-|A|
\choose |G|-i}}{{|E| \choose |G|}}.$ We use the $p$-value threshold of
$0.05$.

\vspace{0.1cm} \hspace{-0.33cm}\textbf{Gene Ontology (GO) enrichment.}
We study the enrichment of a data set in biological process GO terms
\citep{GENEONTOLOGY}. We use: 1) all 4,913 GO terms that annotate
(independent on the evidence code) at least two genes from
StatNetExpression and 2) 2,088 GO terms that annotate (with respect to
an experimental evidence code only) at least two genes from
StatNetExpression. For a GO term $g$, we compute the statistical
significance of its enrichment via the above hypergeometric test
formula, where now $E$ is the set of genes from StatNetExpression that
are annotated by any GO term, $A$ is the gene set in which we are
measuring GO term enrichment, $G$ is the subset of genes from $E$ that
are annotated by GO term $g$, and $O$ is the set of genes in the
overlap between $A$ and $G$.  We use the $p$-value threshold of
$0.05$.

\vspace{0.1cm} \hspace{-0.33cm}\textbf{GO term overlap.} We measure the 
statistical significance of the overlap of GO terms enriched in one
data set and GO terms enriched in another data set via the above
hypergeometric test formula, where now $E$ is the set of GO terms that
annotate at least two genes from StatNetExpression, $A$ is the set of
GO terms enriched in any one of the two data sets, $G$ is the set of
GO terms enriched in the other data set, and $O$ is the set of GO
terms that are in the overlap between $A$ and $G$.  We use the
$p$-value threshold of $0.05$.

\vspace{0.1cm} \hspace{-0.33cm}\textbf{Disease Ontology (DO)
  enrichment.}  We study the enrichment of a data set in all 517 DO
  terms that annotate at least two genes from StatNetExpression
  \citep{Du2009} in the same way as when we study GO term enrichments.

\vspace{0.1cm} \hspace{-0.33cm}\textbf{DO term overlap.}  We study the 
overlap of DO terms from different data sets in the same way as when
we study GO term overlaps.

\vspace{0.1cm} \hspace{-0.33cm}\textbf{Literature validation.}  We 
\emph{automatically} search for a  gene in PubMed 
(http://www.pubmed.gov) and consider the gene to be validated in the
context of aging if its name is mentioned (according to NCBI's
E-utilities -- http://www.ncbi.nlm.nih.gov/books/NBK25500/) with
``age'', ``aging'', or ``ageing'' in the title or abstract of at least
one article. Also, we \emph{manually} search for a gene by reading
relevant PubMed articles \emph{more closely}.

%We aim
%to validate our predictions by searching for them in
%PubMed (http://www.pubmed.gov) that comprises of more than 22
%million of biomedically-related articles. We say that a predicted gene
%is successfully validated if the name of the gene is mentioned
%(according to NCBI's
%E-utilities -- http://www.ncbi.nlm.nih.gov/books/NBK25500/)
%with at least one of the following aging-related keywords: ``age'',
%``aging'', or ``ageing'', in the title or abstract of at least $k$
%articles.  Note that since the title and abstract of an article
%represent only a small portion of the entire article, the validation
%accuracy could actually be higher than what we report.

\end{methods}

%\vspace{-0.1cm}

\section{Results and discussions}

We study \emph{global} topologies of the age-specific networks in
%do
%\emph{not} change with age (
Section \ref{sect:results_global_analysis}.
%We find that this is not the case.
%global properties of the
%networks do not vary with age (Section
%\ref{sect:results_global_properties}), that networks at different ages
%belong to the same graph family (Section
%\ref{sect:results_model_fit}), and that  intersections between the
%networks are  large (Section \ref{sect:intersections}).
%Since global network topologies do not significantly change with age  
%Since this is \emph{not} the case , 
We study \emph{local} topologies of proteins in each network and
predict
%with
%respect to each centrality measure, in order to identify proteins
%whose centralities significantly change with age. We predict such
%proteins as 
aging-related genes in Sections \ref{sect:results_prediction} and
%). 
%Most  of the predicted genes
%are negatively correlated with age -- their centralities decrease as
%age increases. 
%We score our predictions so that the more centrality
%measures support a prediction,
% and the higher the significance of the
%change of centrality values with age, 
%the more credible the prediction (Section
\ref{sect:results_relationship}.
We validate our predictions in Section~\ref{sect:results_validation}.
%Their overlap with ``ground
%truth'' aging-related data is statistically significant. Our
%highest-scoring predictions are present in multiple ``ground truth''
%data sets. Yet, many of our predictions are
%\emph{novel}, confirming that dynamic network analysis of integrated
%data can reveal \emph{additional} biological knowledge. Our
%predictions group by biological processes and diseases that are
%different than those of genes that we do not predict as aging-related.
%Overlap between biological processes and diseases that are enriched in
%our aging-related predictions and those that are enriched in ``ground
%truth'' aging-related data is statistically significant. We give
%evidence that all diseases that are enriched in our predictions are
%linked to aging. We validate in the literature all of our 10\%
%highest-scoring novel predictions.

\vspace{-0.15cm}

\subsection{Global network topologies  do not  change with age}
\label{sect:results_global_analysis}

\subsubsection{Global properties of age-specific networks are similar}
\label{sect:results_global_properties}
Average clustering coefficients, average diameters, and graphlet
frequency distributions (Section
\ref{sect:methods_compare_global}) of the age-specific networks
do not significantly change with age (Supplementary Fig.
\ref{supple:fig:global-hprdall} and \ref{supple:fig:global-biogridall}).

\vspace{-0.2cm}

%\begin{figure}[htbp!]
%\centering
%\includegraphics[scale=0.304]{figures/fig-gdd-agreement}
%\vspace{-0.4cm}
%\caption{Global behaviors of the 30 age-specific networks in terms of
%  the fit of each network to six network models (ER, ERDD, GEO, GEOGD,
%  SF and SFGD) with respect to GDD-agreement. Each network is compared
%  to 10 random network instances of the given model, and the fit is
%  averaged over the 10 instances. Error bars correspond to the
%  standard deviations. The higher the value of the average
%  GDD-agreement, the better the fit. Clearly, the same model fits the
%  best all of the 30 age-specific networks.  Results are consistent
%  when we use BioGRID data as the static PPI network instead of HPRD
%  data (Supplementary Fig.  \ref{supple:fig:global-biogridall}).}
%\label{fig:global}
%\end{figure}

%\vspace{-0.2cm}

\subsubsection{Networks at different ages belong to the same graph family}
\label{sect:results_model_fit}

We compare the fit of the age-specific networks to six network models
(Section~\ref{sect:methods_models}). The best-fitting model does not
change with age (Supplementary Fig. \ref{supple:fig:global-hprdall}
and \ref{supple:fig:global-biogridall}). Note that our primary goal is
not to identify the best-fitting model for dynamic PPI networks.
Nonetheless, consistent to results for static PPI networks
\citep{GeoGD,GraphCrunch2,Ratman2009}, it is gene duplication models
that fit the age-specific networks the best.

\vspace{-0.2cm}

\subsubsection{Overlap of age-specific networks is large}
\label{sect:intersections}

%By measuring the overlap between each pair of age-specific networks,
%we find that 
The age-specific networks share \emph{on average} 93\% of the nodes
and 90\% of the edges, depending on age, while every pair of the
networks shares \emph{at least} 86\% of the nodes and 79\% of the
edges (Supplementary Fig. \ref{supple:fig:network-overlap}). Hence,
the network overlaps are quite large.
%explain why the overall topologies of the
%age-specific networks do not vary with age.

\vspace{-0.15cm}

\begin{figure*}[htb]
\centering
\subfigure[]
{ \includegraphics[scale=0.238]{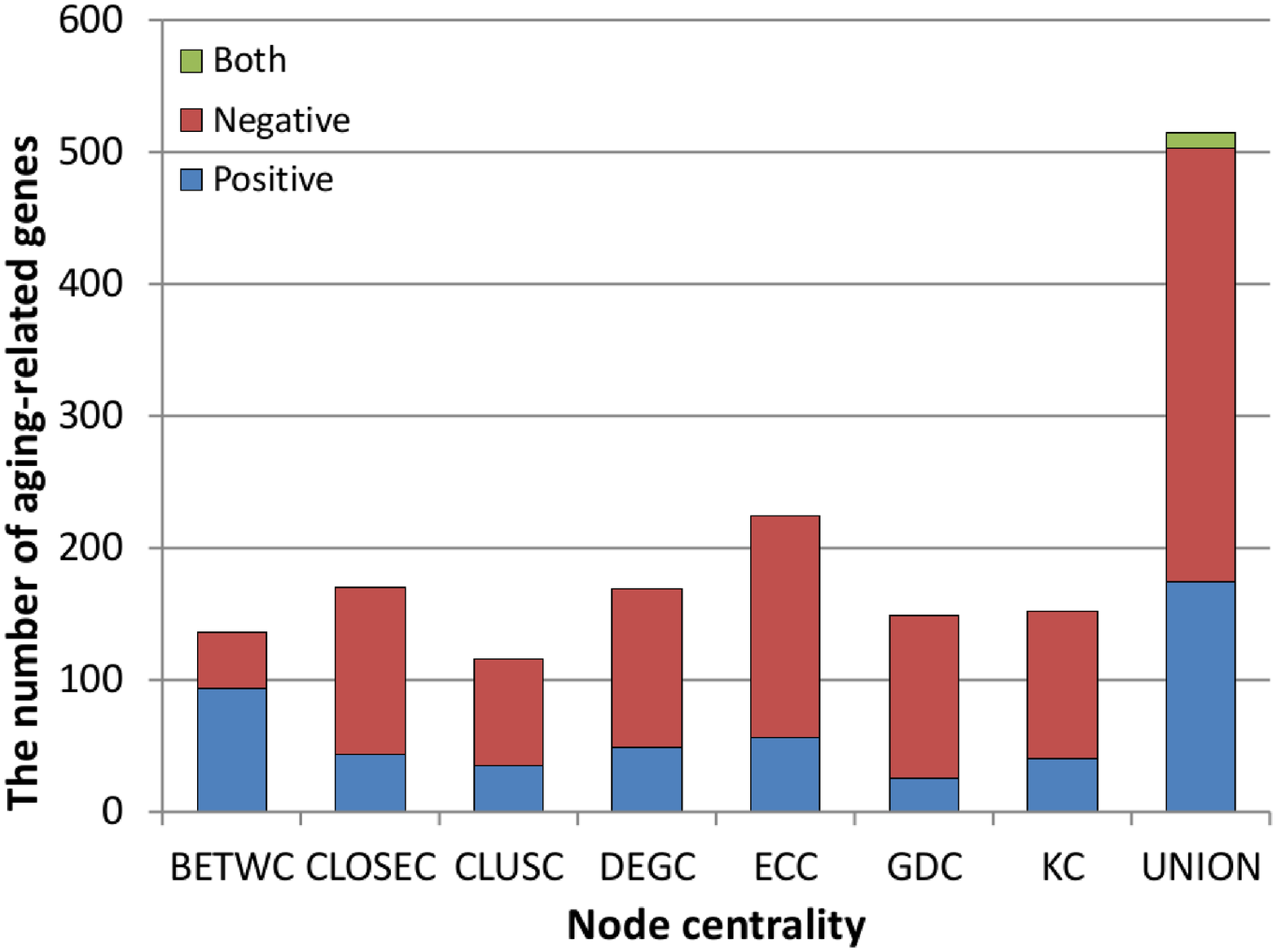} \label{} }
\subfigure[]
{
	\includegraphics[scale=0.238]{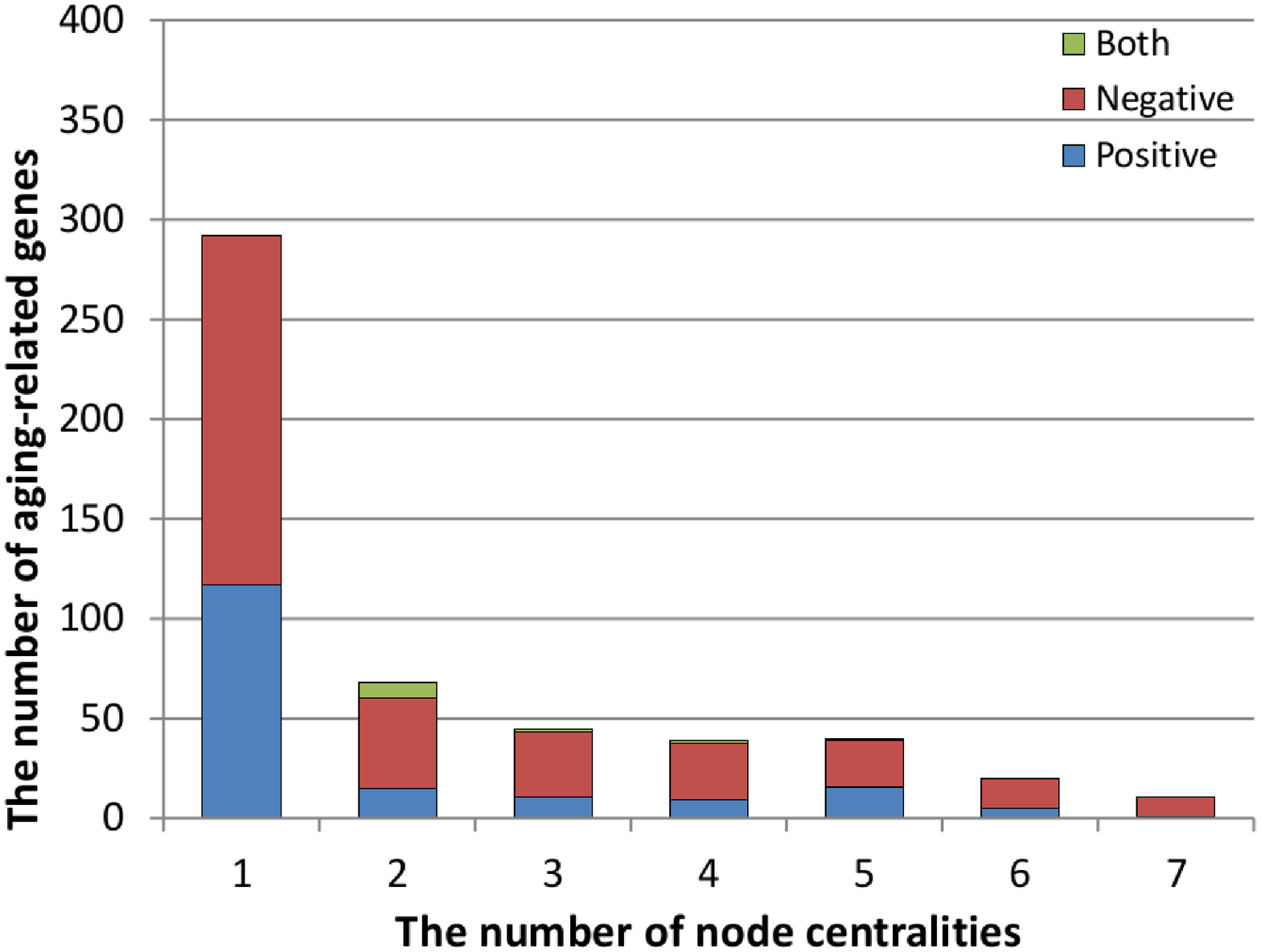}
	\label{}
}
\subfigure[]
{
	\includegraphics[scale=0.238]{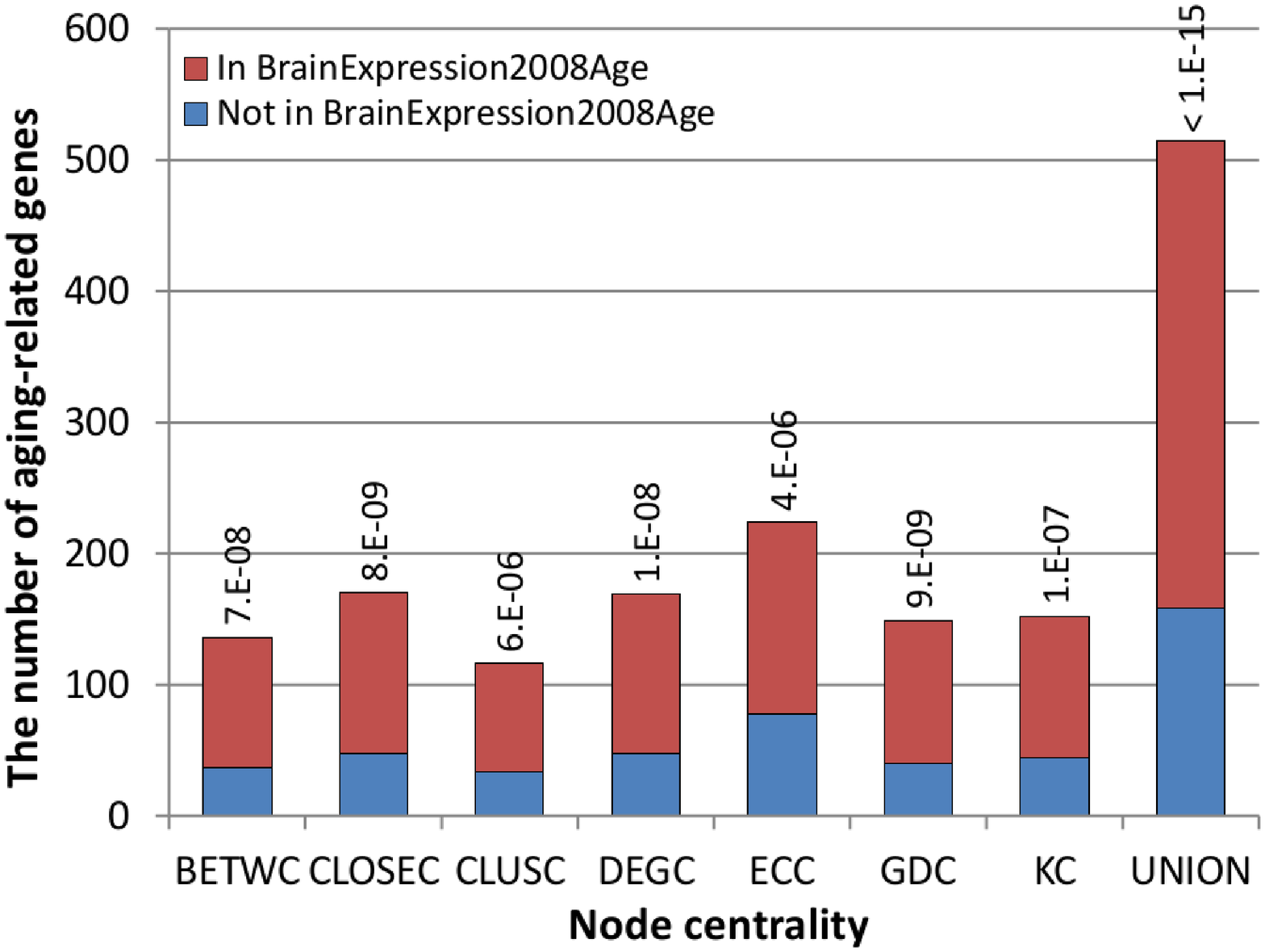}
	\label{}
}
\vspace{-0.45cm}
\caption{The number of our predicted aging-related genes.  Panel
  \textbf{(a)} shows the number of predictions for each of seven node
  centralities individually (BETWC, CLOSEC, CLUSC, DEGC, ECC, GDC and
  KC) or by at least one of them (UNION). Panel \textbf{(b)} shows the
  number of genes predicted by exactly $k$ node centralities ($k = 1,
  2, \dots, 7$).  In the panels, blue and red bars show the number of
  genes that are positively and negatively correlated with age,
  respectively. Green bars denote the number of genes for which one
  centrality measure identifies the given gene as positively
  correlated with age, while another measure identifies the same gene
  as negatively correlated with age.  Panel \textbf{(c)} shows the
  overlaps of our predictions with aging-related genes from
  BrainExpression2008Age ``ground truth'' data, for each centrality
  individually or for all centralities combined. In the panel, blue
  and red bars show the number of predicted genes that are absent from
  and present in the ``ground truth'' data, respectively.  The
  $p$-values for the overlaps are noted at the top of the bars. The
  results are consistent when we use Spearman correlation to predict
  aging-related genes instead of Pearson correlation (Section
  \ref{sect:corr-age-cent} and Supplementary
  Fig. \ref{supple:fig:spearman-all}). Also, the results are
  consistent when we use BioGRID data as the static PPI network
  instead of HPRD data (Section \ref{sect:dynamicppinetworks} and
  Supplementary Fig. \ref{supple:fig:noofcent-all}).}
\label{fig:age-gene-posneg}
\end{figure*}

\subsection{Local topologies of proteins do change with age}
\label{sect:results_local_analysis}

\subsubsection{Prediction of aging-related genes}\label{sect:results_prediction}

Gene expression data alone revealed only 442 out of thousands of genes
as aging-related \citep{Lu2004}.  Thus, while global network analysis
failed to uncover any aging-related information, it could be that the
dynamic network data encodes aging-related information only locally
and around only a subset of nodes. So, we use node centrality measures
(Section~\ref{sect:methods_NC}) to quantify local positions of nodes
in the age-specific networks and find nodes whose centralities
correlate well with age, as such proteins could be key players in
aging.
% and potential drug targets.

We predict a gene as aging-related if its centrality values are
statistically significantly correlated with age (Section
\ref{sect:corr-age-cent}) for  at least one centrality
measure. This results in 515 (8.1\%) predictions out of all 6,397
genes.  Fig. \ref{fig:age-gene-posneg} (a) shows the number of
aging-related predictions for each centrality individually and all
centralities combined.
%Although some centralities predict more genes than others, 
No centrality predicts drastically more genes than others.

A gene's centrality can be positively correlated with age (the gene
becomes more network-central with age) or it can be negatively
correlated with age (the gene becomes less central with age).  The
majority of our predictions are negatively correlated with age
(Fig. \ref{fig:age-gene-posneg}).  This finding is encouraging, since
it has already been argued that aging is associated with failure of
``hubs'' -- highly interconnected and thus network-central proteins
\citep{Soltow2010}.

\vspace{-0.3cm}

\subsubsection{Relationships and potential redundancies of different
  node centralities}\label{sect:results_relationship}

%Different centralities capture different aspects of the network
%position of a node. So, we aim to study their relationships and
%potential redundancies.

We predict a gene as aging-related if its centrality values correlate
well with age with respect to \emph{at least one} centrality. So, we
study whether any genes are predicted by more than one or even all of
the centralities.  We find that almost half (43\%) of the 515
aging-related predictions are supported by multiple centralities,
while the remaining predictions are supported by a single centrality
(Fig.~\ref{fig:age-gene-posneg} (b)). As expected, the number of
predictions decreases as the number of centralities supporting the
predictions increases.

\begin{figure}[htb]
\centering
\hspace{-0.1cm}
\subfigure[]
{ 
	\hspace{-0.4cm}
	\includegraphics[scale=0.305]{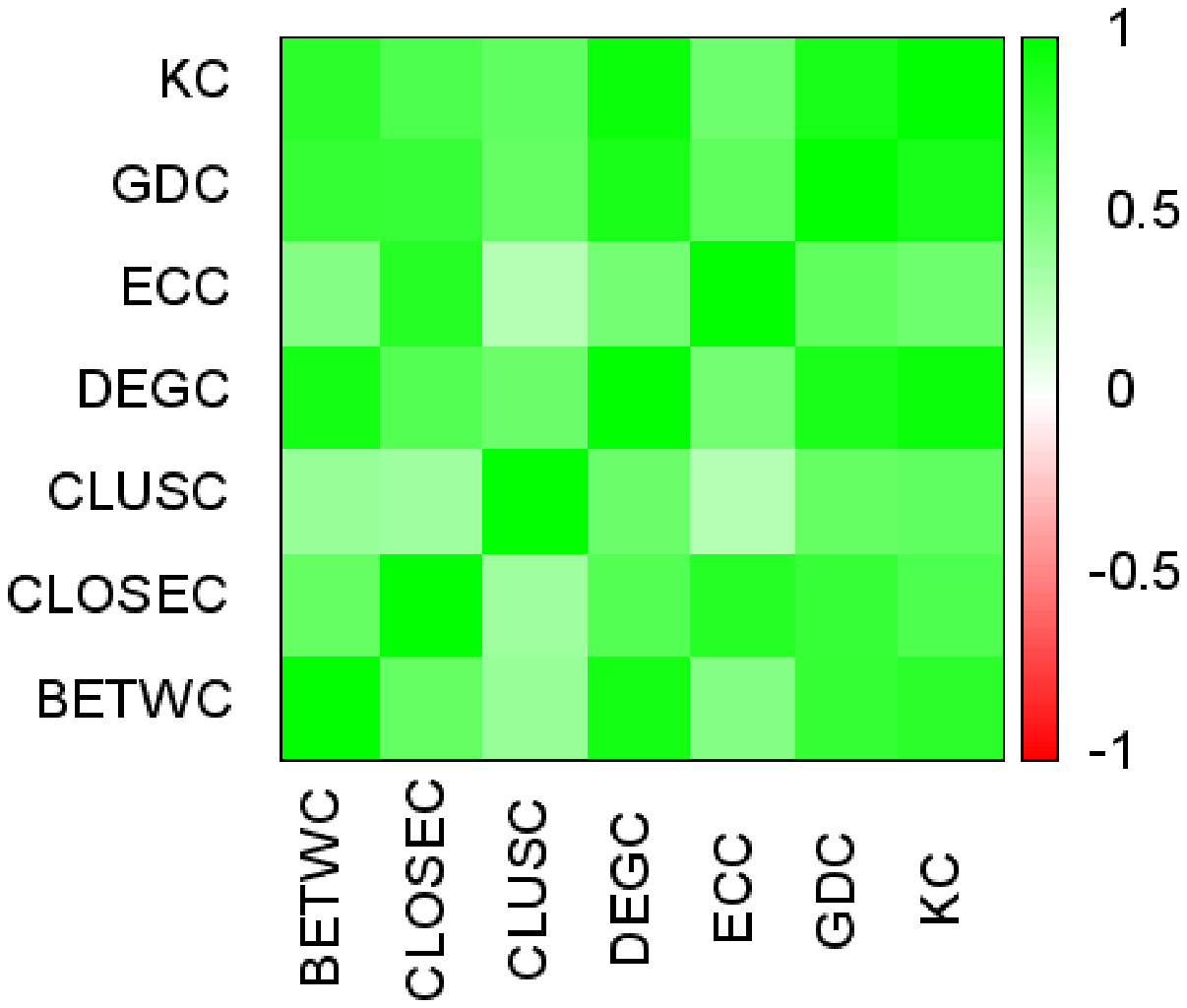} 
	\label{} 
}
\hspace{0.5cm}
\subfigure[]
{
	\hspace{-0.4cm}
	\includegraphics[scale=0.305]{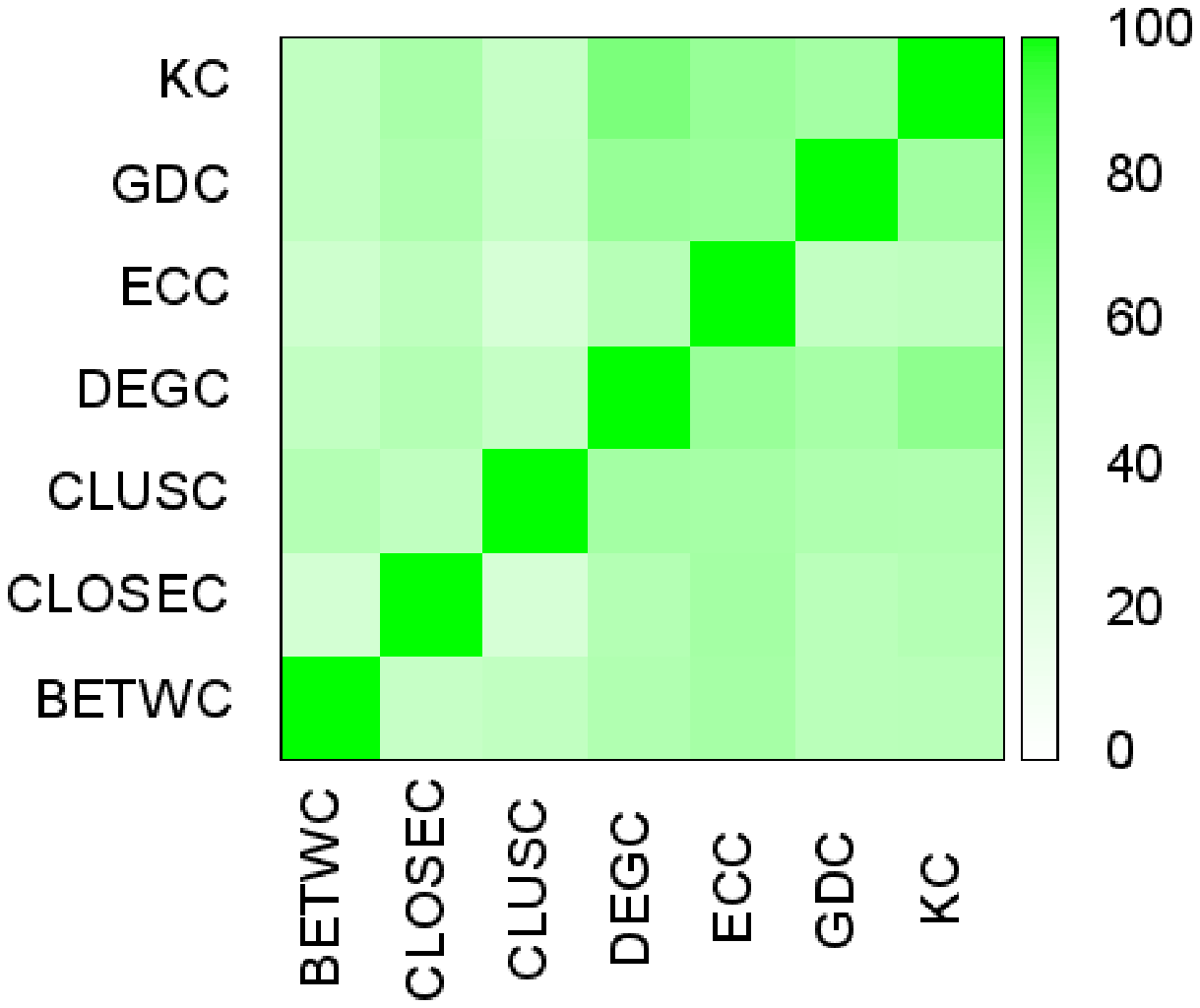}
	\label{}
}\\
\hspace{0.2cm}
\subfigure[]
{
	\hspace{-0.4cm}
	\includegraphics[scale=0.305]{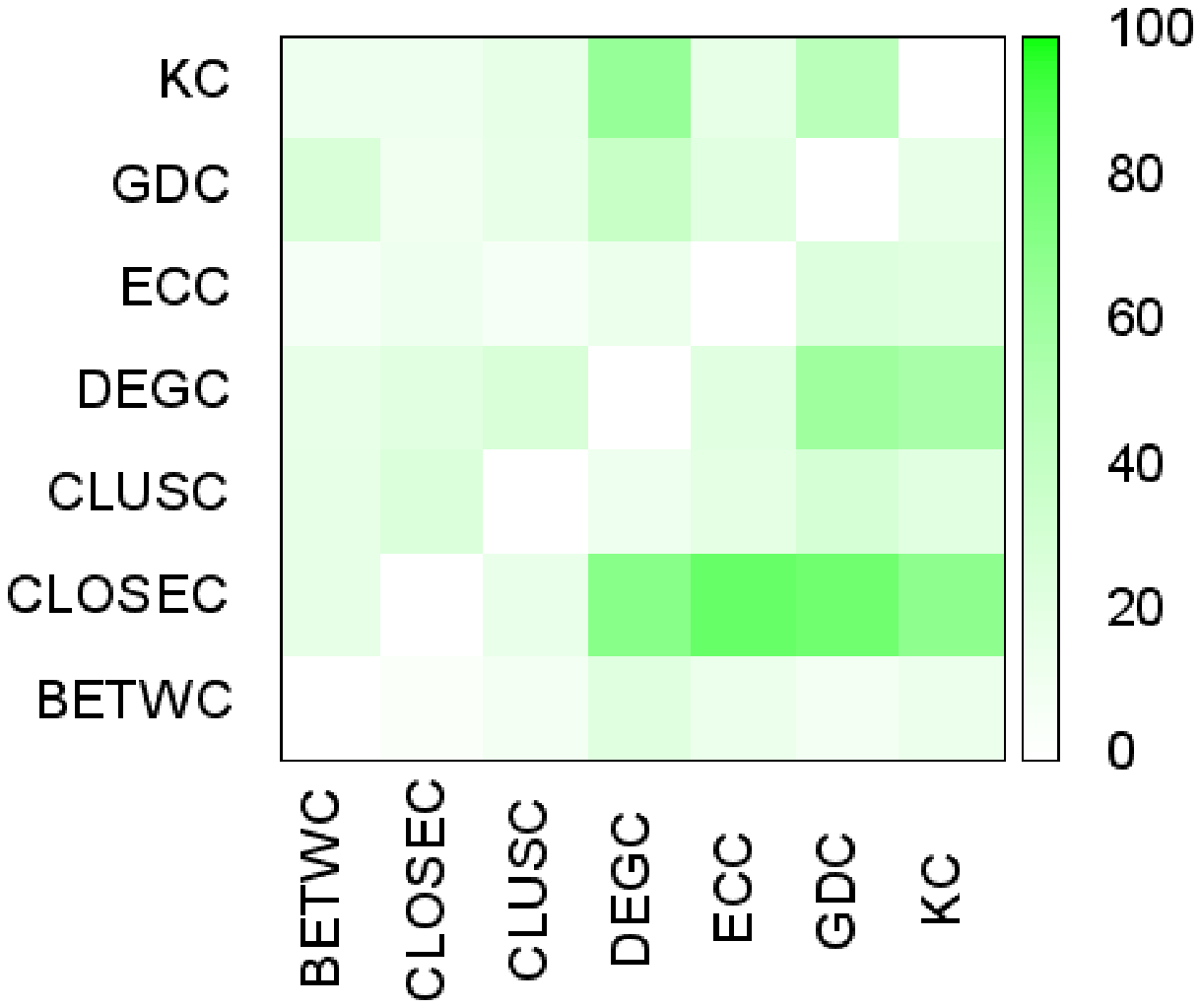}
	\label{}
}
\hspace{0.2cm}
\subfigure[]
{
	\hspace{-0.4cm}
	\includegraphics[scale=0.26]{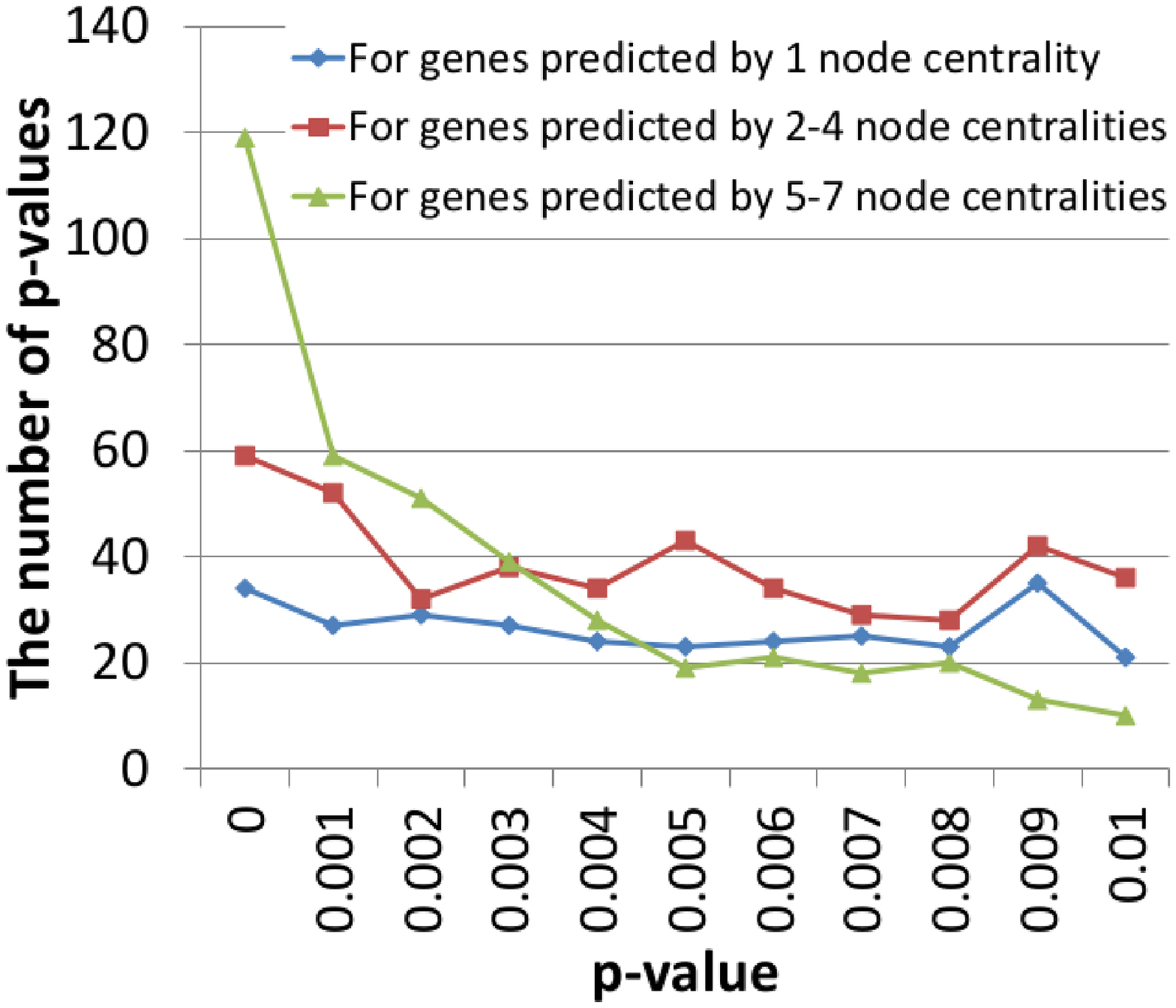}
	\label{}
}
\vspace{-0.6cm}
\caption{Relationships between different centralities (BETWC, CLOSEC,
  CLUSC, DEGC, ECC, GDC and KC): \textbf{(a)} Spearman correlation
  between each pair of centralities averaged over the 30 age-specific
  networks; \textbf{(b)} pairwise overlap of aging-related genes
  predicted by the different centralities; \textbf{(c)} percentage of
  genes predicted as aging-related ($p$-value $\le $ 0.01) by exactly
  one centrality (listed in a given row), which have a ``marginal''
  $p$-value between 0.01 and 0.05 with respect to one of the six
  remaining centralities (listed in a given column), and \textbf{(d)}
  distributions of $p$-values of our predicted aging-related
  genes. Panel (c) can be interpreted as follows. For example, many of
  the predictions identified only by CLOSEC (the second last row) are
  marginally significant with respect to DEGC, ECC, GDC, and KC
  (intensive color), while almost none of the predictions identified
  only by BETWC (the last row) are marginally significant with respect
  to any other centrality (light color). Hence, the predictions made
  by CLOSEC only may be more credible than the predictions made by
  BETWC only, as the former are marginally supported by additional
  centralities, but the latter are not.  }
\label{fig:heatmap-centvscent}
\vspace{-0.4cm}
\end{figure}

We study the redundancy of the different centralities by computing,
for each pair of centralities, the correlation between their
centrality values over all nodes in the given network, and by
averaging correlations over the 30 age-specific networks. We observe
high correlations between some measures, such as DEGC, KC, and GDC, or
CLOSEC and ECC (Fig.~\ref{fig:heatmap-centvscent} (a) and
Supplementary Section
\ref{supple:supplesect:results-redundancy}).  Thus, some centralities appear to be redundant to others.

However, when we study pairwise overlaps of aging-related predictions
produced by the different centralities, the overlaps are not very
large for any pair of centralities (Fig.~\ref{fig:heatmap-centvscent}
(b)).  This result, together with the result from
Fig.~\ref{fig:age-gene-posneg} (b), which shows that the majority
($\sim$57\%) of the predictions are identified by a single centrality,
suggests that the centralities are not redundant to each other.  When
we focus on our aging-related genes predicted by exactly one
centrality measure, in most cases, these predictions are not even
marginally significant with respect to other centrality measures (Fig.
\ref{fig:heatmap-centvscent}(c) and Supplementary Fig. 
\ref{supple:fig:heatmap-centvscent}), indicating again that the different
centralities are in general not redundant to each other.  Thus, we
keep all 515 predictions, independent on the number of centralities
supporting the given prediction. Henceforth, we denote this set of 515
aging-related genes predicted via dynamic network analysis as
\emph{DyNetAge}. We denote the complement of this set, i.e., the
set of genes that are present in StatNetExpression but not in
DyNetAge, as \emph{DyNetComplement}. Clearly, DyNetComplement is the
set of genes whose centralities do not significantly correlate with
age, and as such, we do not predict them as aging-related.

Next, we study the effect of the number of centralities supporting a
prediction on the quality of the prediction. It is not necessarily the
case that predictions supported by many centralities are more enriched
in ``ground truth'' aging-related genes (see below) compared to
predictions supported by only few centralities (Supplementary Fig.
\ref{supple:fig:overlap-noofcent-gt-all}). Nonetheless, it \emph{is}
the case that the more centralities support a prediction, the more
significant the correlation of its centrality values with age (as
indicated by lower $p$-values in
Fig. \ref{fig:heatmap-centvscent}(d)). Hence, to account for the
number of centralities supporting a prediction, we rank the prediction
so that the more centralities support it and the more significant the
change of its centrality values with age, the more credible the
prediction (Section~\ref{sect:corr-age-cent}).  (Supplementary Tables
\ref{supple:table:dynetage} and
\ref{supple:table:dynetagebiogrid} contain the ranked lists of all
predictions.) We validate our scoring scheme by demonstrating that our
10\% highest-scoring predictions are statistically significantly
enriched in genes that are present in more than one ``ground truth''
data set ($p$-value of 0.003), whereas this is not the case for
the lower-scoring predictions.

%\vspace{-0.4cm}

\vspace{-0.25cm}

\subsubsection{Validation of predicted aging-related
  genes}\label{sect:results_validation}

\emph{}

\vspace{0.1cm} \hspace{-0.35cm}\textbf{Dynamic network analysis gives
  meaningful and statistically significant aging-related predictions.}
The number of aging-related genes in DyNetAge is statistically
significantly larger than the number of aging-related genes that we
can predict from the ``randomized'' data ($Z$-score$=7.42$, $p$-value
$< 10^{-10}$; Section~\ref{sect:methods_validation}). 
%This holds for
%the 515 predictions combined across all centralities as well as for
%predictions produced by each centrality individually.

\vspace{0.1cm} \hspace{-0.35cm}\textbf{Overlap of our predictions with
  ``ground truth'' aging-related genes is statistically significant.}
  Human aging is hard to study experimentally due to long life span
  and ethical constraints. Instead, human aging-related ``ground
  truth'' knowledge has been \emph{predicted} more or less
\emph{computationally}, by studying gene expression data or by
transferring aging-related knowledge from model species to human via
sequence comparison. (We intentionally use quotes, as we are
\emph{not} dealing with true, experimentally obtained ground truth
data.) Similarly, here, we aim to computationally predict new ``ground
truth'' aging-related data from an additional data type -- PPIs. (But
by no means do we claim to identify \emph{all} aging-related genes.)
Hence, since ``ground truth'' data sets are predicted computationally,
they could be noisy.  Also, different ``ground truth'' sets could be
biased towards different data types from which the predictions have
been made, be it expression, sequence, or PPI data. Since different
data types could be capturing different functional slices of the cell
\citep{Memisevic10b,Przytycka2010},
%due to potential complimentarity of different ``ground truth'' data sets,
it might not be alarming if the intersections between the different
``ground truth'' sets are not very large. However, since all the sets
are aiming to capture the same biological phenomenon (aging), some
overlap would be encouraging. Ideally, we would like to see a
statistically significant overlap. However, the existence of some
overlap would be encouraging even if the overlap was not significant,
since: (1) the overlap could be low due to the noisiness of each of
the ``ground truth'' sets, and (2) statistically non-significant
results may be biologically important, whereas statistically
significant results may not be
\citep{Motulsky1995,MMGP_Roy_Soc_09,Ho2010}.

We measure the overlap of DyNetAge with five ``ground truth'' data
sets: BrainExpression2004Age, BrainExpression2008Age, ADExpressionAge,
HGPSExpressionAge, and SequenceAge (Section \ref{sect:ground_truth}).
Of the five, BrainExpression2004Age and BrainExpression2008Age are the
most likely to be similar to DyNetAge (as all three are based on
\emph{brain-related} data, brain-related \emph{gene expression} data,
and brain \emph{aging-related} gene expression data; Supplementary
Section \ref{supple:supplesect:results-overlap-gt}), followed by
ADExpressionAge (as our predictions as well as this data are both
based on \emph{brain-related} data and brain-related \emph{gene
expression} data), followed by HGPSExpressionAge (as our predictions
as well as this data are both based on \emph{gene expression} data),
followed by SequenceAge (as our predictions and this data both capture
aging-related information but from different data types) (Section
\ref{sect:ground_truth}). Therefore, a high overlap of DyNetAge with
BrainExpression2004Age or BrainExpression2008Age would validate our
method. A high overlap with ADExpressionAge would suggest that our
method could capture not only brain aging-related genes but also brain
aging-related \emph{disease} genes. A high overlap with
HGPSExpressionAge would suggest that our method could capture genes
related to a tissue different from the brain tissue.  A high overlap
with SequenceAge would suggest that our method could capture genes
identified from a different data type, namely sequence data.
 
%One might argue that because DyNetAge and BrainExpression2004Age are
%(partly) based on the same data set, the existence of a high overlap
%is expected. We argue that this is
%\emph{not} the case (Supplementary Section \ref{XXX}). Nonetheless, 
%to avoid a potential circular argument, we use BrainExpression2008Age
%(and the three other data sets) as additional ``ground truth''
%aging-related data, because BrainExpression2008Age closely resembles
%BrainExpression2004Age but represents an independent data set that we
%did not use when producing our DyNetAge predictions.  And actually, we
%show that it is not necessarily the case that DyNetAge has a higher
%overlap with BrainExpression2004Age than with BrainExpression2008Age
%(see Supplementary Fig.
%\ref{supple:fig:overlap-gt-all} and below).

As hypothesized, DyNetAge overlaps the most (as indicated by the
lowest $p$-values) with BrainExpression2004Age and
BrainExpression2008Age, followed by ADExpressionAge,
HGPSExpressionAge, and SequenceAge, respectively (Supplementary Fig.
\ref{supple:fig:overlap-gt-all} and Supplementary Table \ref{supple:tab:overlap-gt}). The overlap is statistically
significant for BrainExpression2004Age, BrainExpression2008Age, and
ADExpressionAge, marginally significant for HGPSExpressionAge, and
non-significant for SequenceAge (Table \ref{table:all-overlap-pv}).
The (marginally) significant overlap between DyNetAge and four out of
the five ``ground truth'' data sets is encouraging.  Importantly, even
though most of the overlaps are statistically significant, 87\%, 31\%,
74\%, 87\%, and 96\% of our DyNetAge predictions are not in
BrainExpression2004Age, BrainExpression2008Age, ADExpressionAge,
HGPSExpressionAge, and SequenceAge, respectively, and 19\% of our
predictions are not in \emph{any} of the five data sets (Supplementary
Table \ref{supple:table:novel-nonnovel}).  This confirms that data
integration can reveal \emph{additional} biological knowledge compared
to studying individual data types.
%, e.g., gene expression data alone.

Some overlap between DyNetAge and SequenceAge is also encouraging,
even though the overlap is non-significant (as argued above).
Further, the overlap is \emph{stronger} between DyNetAge and the other
four ``ground truth'' data sets (BrainExpression2004Age,
BrainExpression2008Age, ADExpressionAge, and HGPSExpressionAge) than
between SequenceAge and these four ``ground truth'' data sets. In
particular, whereas DyNetAge overlaps significantly with three of the
four data sets and marginally significantly with respect to the fourth
data set (see above), SequenceAge overlaps significantly with just one
of the four data sets (ADExpressionAge), almost marginally
significantly with respect to two of the four data sets
(HGPSExpressionAge and BrainExpression2004Age), and non-significantly
with respect to the remaining data set (BrainExpression2008Age)
(Supplementary Table \ref{supple:tab:overlap-gt}).  Therefore, our
DyNetAge appears to be more relevant than SequenceAge with respect to
the other ``ground truth'' data sets.  The non-significant overlaps
could be due to potential complementarity of the different types of
biological data, noisiness of the ``ground truth'' data, or some of
the ``ground truth'' data sets being biased towards
\emph{brain}-related genes.

Recall that the complement of each ``ground truth'' data set
(including DyNetAge) is the set of genes not predicted as
aging-related by the given study (Section
\ref{sect:complements_ground_truth}). Hence, it would be encouraging
to see: \textbf{1)} low (non-significant) overlaps between DyNetAge
and complements of the ``ground truth'' data sets,
\textbf{2)} low (non-significant) overlaps between DyNetComplement and
``ground truth'' sets, and \textbf{3)} high (significant) overlaps
between DyNetComplement and complements of ``ground truth''
sets. Indeed, this is what we typically observe in all three cases
(Supplementary Table
\ref{supple:tab:overlap-gt}).

\newcolumntype{A}{>{\centering\arraybackslash}m{0.68cm}}
\newcolumntype{B}{>{\centering\arraybackslash}m{0.98cm}}
\newcolumntype{C}{>{\centering\arraybackslash}m{1.03cm}}
\newcolumntype{D}{>{\centering\arraybackslash}m{0.86cm}}
\newcolumntype{E}{>{\centering\arraybackslash}m{0.70cm}}

\vspace{-0.2cm}
\begin{table}[htbp!]
\centering
\caption{Overlap and its statistical significance (i.e. $p$-value) between aging-related genes (ARG), GO terms  (GO), and DO terms (DO) in DyNetAge and those in the five ``ground truth'' aging-related data sets (BrainExpression2004Age (BE4A), BrainExpression2008Age (BE8A), ADExpressionAge (ADEA), HGPSExpressionAge (HEA), and SequenceAge (SA)).}
\vspace{-0.25cm}
\begin{tabular}{|A|B|C|C|D|E|E|}\hline
{} & {} & BE4A & BE8A & ADEA & HEA & SA \\\hline ARG & Overlap & 20\%
& 69\% & 26\% & 13\% & 9\% \\ {} & $p$-value & 5.5E-13 & $<$1E-15 &
4.3E-7 & 0.06 & 0.39 \\
\hline
GO & Overlap & 20\% & 21\% & 12\% & 6\% & 17\% \\
{} & $p$-value & 1.2E-10 & 2.5E-14 & 2.8E-6 & 0.28 & 0.33 \\
\hline
%GO & Overlap & 5\% & 14\% & 0\% & 0\% & 2\% \\
%{exp} & $p$-value & 0.23 & 6.9E-4 & N/A & N/A & N/A \\
%\hline
DO & Overlap & 25\% & 25\% & 0\% & 0\% & 25\% \\
{} & $p$-value & 0.03 & 7.2E-3 & N/A & N/A & 0.70 \\
\hline
\end{tabular}\label{table:all-overlap-pv}
\end{table}

\newcolumntype{A}{>{\centering\arraybackslash}m{1.75cm}}
\newcolumntype{B}{>{\centering\arraybackslash}m{0.70cm}}

\vspace{-0.3cm}

\vspace{0.1cm} 

\hspace{-0.35cm}\textbf{GO enrichment.} When analyzing all gene-GO
term associations, 146 GO terms are significantly enriched in DyNetAge
($p$-values between $0.047$ and $1.5 \times 10^{-5}$), while only 14
GO terms are enriched in DyNetComplement ($p$-values between $0.043$
and $5.1 \times 10^{-5}$) (Section \ref{sect:methods_validation}).
Importantly, there is no overlap between the GO terms from DyNetAge
and those from DyNetComplement.  Hence, our aging-related predictions
group by functions that are different than functions of genes that we
do not predict as aging-related.

When we focus on \emph{aging-related} GO terms only, it is encouraging
that DyNetAge contains genes annotated with \emph{aging},
\emph{cell aging}, and \emph{cellular senescence} GO terms, as well as that  
DyNetAge's performance is typically comparable to that of the ``ground
truth'' data, especially when using only gene-GO term associations
obtained by
\emph{experimental} evidence codes (Supplementary Section
\ref{supple:supplesect:results-go}).

\vspace{0.1cm} 

\hspace{-0.35cm}\textbf{GO overlap.} Our predictions are further 
validated by: \textbf{1)} high overlaps between GO terms from DyNetAge
and GO terms from the ``ground truth'' data sets,
\textbf{2)} low overlaps between GO terms from DyNetAge and GO terms
from complements of the ``ground truth'' sets, \textbf{3)} low
overlaps between GO terms from DyNetComplement and GO terms from the
``ground truth'' sets, and \textbf{4)} high overlaps between GO terms
from DyNetComplement and GO terms from complements of the ``ground
truth'' sets (Table \ref{table:all-overlap-pv} and Supplementary Table
\ref{supple:tab:overlap-go-all}).  For example, when considering
gene-GO term associations of any evidence code, GO terms from DyNetAge
significantly overlap with GO terms from three of the five ``ground
truth'' data sets (BrainExpression2004Age, BrainExpression2008Age, and
ADExpressionAge), and some (though non-significant) overlap with GO
terms from HGPSExpressionAge and SequenceAge is also encouraging
(Table
\ref{table:all-overlap-pv}).
%For other results, see Supplementary Section
%\ref{supple:supplesect:results-go-overlap}.
%
Equivalent results when considering gene-GO term associations of
\emph{experimental} evidence codes \emph{only} are shown in Supplementary Table
\ref{supple:tab:overlap-go-exp}. Importantly, in this case, SequenceAge fails 
to significantly overlap with any other ``ground truth'' set, whereas
DyNetAge still significantly overlaps with BrainExpression2008Age.
For details, see Supplementary Section
\ref{supple:supplesect:results-go-overlap}.

\vspace{0.1cm} 

\hspace{-0.35cm}\textbf{DO enrichment.}  Eight diseases are
significantly enriched in DyNetAge ($p$-values between $0.041$ and $6
\times 10^{-3}$), while 14 diseases are enriched in DyNetComplement
($p$-values between $0.049$ and $2.9 \times 10^{-3}$) (Section
\ref{sect:methods_validation}).  Importantly, there is no overlap
between diseases from DyNetAge and those from DyNetComplement.

The eight diseases in DyNetAge are: brain disease, brain tumor,
bipolar disorder, connective tissue disease, renal tubular acidosis,
leukodystrophy, neuroblastoma, and demyelinating disease.  \emph{Brain
disease} and \emph{renal tubular acidosis} are enriched in SequenceAge
as well, \emph{brain disease} and \emph{bipolar disorder} are enriched
in BrainExpression2004Age as well, and \emph{brain tumor} and
\emph{neuroblastoma} are enriched in BrainExpression2008Age as
well.  All of these overlaps are encouraging. (We quantify the
significance of the overlaps in the following section.)  In
particular, \emph{renal tubular acidosis}, whose aging-related
evidence is supported by SequenceAge, is kidney-related, whereas
DyNetAge has been predicted from the brain-related data. Capturing
this non-brain-related disease in DyNetAge is encouraging, especially
because BrainExpression2004Age and BrainExpression2008Age fail to do
so. Further aging-related evidence for these five diseases can be
found in the following references denoted by their PubMed IDs (PMIDs):
11256685, 8040891, 21197651, and 21031036.  Supplementary Table
\ref{supple:table:pmid-reference} maps PMIDs to full paper references. 
%\emph{Brain disease} being enriched in two  ``ground truth'' sets,  
%as well as \emph{bipolar disorder}, a brain disease with a risk factor
%for many aging-related diseases, e.g., Parkinson's, Alzheimer's, or
%Huntington's disease (PubMed ID (PMID): 11256685), being enriched in
%BrainExpression2004Age, is encouraging.  Interestingly, Further, it
%encouraging that \emph{brain tumor} and
%\emph{neuroblastoma} (a common extracranial tumor at the age of
%human infancy (PMID: 8040891)) are also enriched in
%BrainExpression2008Age, as cancer and tumor occurrence is directly
%correlated with age (PMIDs: 21197651 and 21031036).
%
Importantly, even three diseases from DyNetAge that are missed by
\emph{all} ``ground truth'' sets can \emph{all} be linked to aging in
the literature as well (Supplementary Section
\ref{supple:supplesect:results-do}).

\vspace{0.1cm}

\hspace{-0.35cm}\textbf{DO overlap.}  
%Similar to GO overlap analysis, 
We further validate our predictions by demonstrating: \textbf{1)} high
overlaps between DO terms from DyNetAge and DO terms from the ``ground
truth'' data sets,
\textbf{2)} low overlaps between DO terms from DyNetAge and DO terms
from complements of the ``ground truth'' sets,
\textbf{3)} low overlaps between DO terms from DyNetComplement and DO
terms from the ``groung truth'' sets, and \textbf{4)} high overlaps
between DO terms from DyNetComplement and DO terms from complements of
the ``ground truth'' sets (Supplementary Table
\ref{supple:tab:overlap-do}). For example, DO terms from DyNetAge
statistically significantly overlap with DO terms from two of the five
``ground truth'' data sets (BrainExpression2004Age and
BrainExpression2008Age) (Table
\ref{table:all-overlap-pv}). It is not alarming that DyNetAge does not
significantly overlap with ADExpressionAge, HGPSExpressionAge, or
SequenceAge, since none of the three overlaps significantly with more
than one of the five ``ground truth'' data sets (Supplementary Table
\ref{supple:tab:overlap-do}). Hence, DyNetAge is better supported by
the ``ground truth'' data than any of these three data sets. For
details, see Supplementary Sections
\ref{supple:supplesect:results-do-overlap} and \ref{supple:supplesect:results-do-nonaging}.

\vspace{0.1cm} \hspace{-0.35cm}\textbf{Literature validation.} 
Automatic literature validation (Section
\ref{sect:methods_validation}) is prone to errors: we 
``validate'' in this manner equal portion of both the ``ground truth''
aging-related sets \emph{and} their complements.
%, for \emph{each} of the ``ground truth'' sets.  
Therefore, we aim to validate our predictions manually (Section
\ref{sect:methods_validation}). Since manual validation is 
laborious, we focus on our 10\% highest-scoring predictions (Section
\ref{sect:corr-age-cent}). Of these, we study predictions that
are absent from all ``ground truth'' sets, namely DVL1, ACACA, HOMER3,
GJB1, FKBP8, and H1F0. We successfully validate
\emph{all} of these genes. DVL1 has been linked to Alzheimer's disease
(PMID: 11803455), which has been linked to aging itself (PMID:
21197651, 21031036). Expression level of ACACA in rat changes with age
(PMID: 11044254), and so its human ortholog is an aging-related
candidate. HOMER-1A, a member of the same family as HOMER3, affects
the level of cognitive performance during aging (PMID: 23054826). The
expression of GJB1 is down-regulated with aging (PMID: 22337502).
Drug targeting of FKBP38 might successfully intervene with
FKBP38-dependent processes such as programmed cell death in cancer or
neurodegenerative diseases (PMID: 21514222), which have been linked to
aging  (PMID: 21197651, 21031036).
H1F0 has been linked to age-related macular degeneration (PMID:
16518403), and its rat ortholog causes aging-related alterations in
liver (PMID: 8114518).

%We
%search for our predictions in PubMed, successfully validating 76\%,
%37\%, and 17\% of the predictions in at least one, 10, and 100
%articles, respectively (Section \ref{sect:methods_validation}).  Also,
%since automatic text mining is prone to errors, we validate our
%top-scoring predictions via \emph{manual} literature search. Of the
%top 10 predictions, we focus on those found in at least five articles
%by automatic text mining and are absent from both ``ground truth''
%data sets. There are five such genes: TIAM1, SYN1, GABRA1, CXCR4 and
%GABRG2, \emph{all} of which we validate, as follows.  It is well known
%that cancer incidence increases with aging, and TIAM1 promotes tumor
%invasion and metastasisis in aging tissues \citep{Liu2012}.  The
%expression level of SYN1 decreases with age and promotes the decline
%in cognitive ability, particularly memory formation
%\citep{Haberman2012}.  GABRA1 is gamma-aminobutyric acid (GABA) A
%receptor, alpha 1.  The responsiveness of GABAergic neurons to
%GABA(A)ergic drugs changes with age, indicating the role of GABRA1 in
%aging \citep{Chudomel2009}. A significant increase in the surface
%expression of CXCR4 in CD4(+) T cells from elderly human donors,
%relative to those from the young, has been observed, and such altered
%regulation of CXCR4 expression during aging contributes to increased
%CXCL12-dependent chemotactic migration of the T cells
%\citep{Cane2012}. Finally, anxiety-related gene GABRG2 has reduced
%expression at old age \citep{Kurz2010}.

\vspace{-0.7cm}

\section{Conclusion}

Together, our results confirm that dynamic PPI network analysis via
integration of static PPI network data with aging-related gene
expression data can reveal meaningful key players in aging.

\vspace{-0.1cm}

\ifx

\section{Response to reviewers' comments:}

\textbf{Comment: The ``ground truth'' gene set BrainExpression2004Age comes from the same data that the authors use to derive their own list of aging-related genes. Their finding of significant overlap between BrainExpression2004Age and DyNetAge would be more convincing if BrainExpression2004Age were derived from an independent dataset. It should be straightforward to download an independent list of genes that change expression with age in human brains from other microarray studies (e.g. Berchtold et al. 2008) and determine if the overlap is still significant.}

\textbf{Response:} Though BrainExpression2004Age and DyNetAge are computed by methods that use the same microarray data as input, a significant overlap between BrainExpression2004Age and DyNetAge can not be assumed due to distinguished nature of corresponding methods. In particular, genes in BrainExpression2004Age are determined based on the significant change in expression levels with age. This method is independent of any information about their PPI network neighbors. On the other hand, genes in DyNetAge are determined based on the significant change in node centralities with age. There are several reasons why the later method is different from the former one. First, the later method uses microarray data only to determine whether a gene is expressed, i.e. the gene is present in an age-specific network at a given age. Second, unlike the former method, the later one incorporates information about PPI network neighbors to determine node centrality of a gene and correlate this information with age to identify whether the gene is aging-related. It is possible that a gene is in BrainExpression2004Age because its expression levels significantly change with age, but the gene is not in DyNetAge because its node centralities do not change significantly with age. Similarly, a gene can be in DyNetAge, but not in BrainExpression2004Age. However, we still measure overlap of DyNetAge with some independent gene expression data to strengthen our validation. To resolve ambiguity, we denote aging-related set of genes predicted from brain gene expression data by \cite{Lu2004} as BrainExpression2004Age.

We introduce three independent gene expression-based ``ground truth'' datasets and measure overlap of DyNetAge with each dataset. The first datasets consists of 8,277 genes that have significantly altered expression levels with age in brain tissue. Among these genes, 3,228 are present in static PPI network. We denote this aging-related set of 3,228 genes as BrainExpression2008Age. Essentially, both BrainExpression2004Age and BrainExpression2008Age are brain aging-related set of genes coming from two independent microarray studies. The second dataset consists of 2,911 genes that are differentially expressed across different stages of Alzheimer's disease (AD). Though these genes are not directly aging-related, we choose this dataset because AD is a well known aging-related brain disease and a significant overlap of DyNetAge with AD-related genes would be encouraging. Among the genes of this dataset, 1,117 are present in static PPI network. We denote this AD-related set of 1,117 genes as ADExpressionAge. The third dataset consists of 1,731 genes that are found within differentially methylated region between Hutchinson-Gilford progeria syndrome (HGPS)-affected and control vascular muscle cells. It is to be noted that HGPS is a fatal human premature aging disease. A significant overlap of DyNetAge with HGPS-related genes would be encouraging in the ability of a dynamic PPI network to capture non-brain aging-related information. Among the genes of this dataset, 708 genes are found in static PPI network. We denote this HGPS-related set of 708 genes as HGPSExpressionAge. 

We find that DyNetAge statistically significantly overlaps with BrainExpression2008Age (p-value of $< 10^{-15}$), which is an independent brain aging-related dataset. DyNetAge statistically significantly overlaps with ADExpressionAge (p-value of $4.26 \times 10^{-7}$) indicating that DyNetAge is capable of capturing brain aging-related disease information. DyNetAge marginally statistically significantly overlaps with HGPSExpressionAge (p-value of 0.06). A significant overlap between DyNetAge and HGPSExpressionAge is not expected because these datasets are originated from the microarray study on two different tissues. It is still encouraging that DyNetAge can marginally capture aging-related information from a different tissue. Interestingly, SequenceAge does not statistically significantly overlap with BrainExpression2008Age (p-value of 0.10). In this particular case, genome sequence alignment fails to capture the aging-related information revealed by microarray study. 

We define the complement of an aging-related dataset to be the set of genes that appear in static PPI network but not in the aging-related dataset. Essentially, the genes in the complement of an aging-related dataset are the ones that are not identified to be aging-related by the dataset. It is expected that most of the genes in the complement dataset is truly non-aging-related. But it is possible that the complement dataset can contain some genes that are identified as aging-related in some other datasets. We measure the overlap of DyNetAge with the complement of each ``ground truth'' aging-related dataset and find that overlaps are not statistically significant (p-values between 0.70 and 1). Therefore, DyNetAge does not have significant overlap with non-aging-related datasets. We also find that the complement of DyNetAge does not have significant overlap with ``ground truth'' aging-related datasets. Therefore, genes that are not identified as aging-related by DyNetAge do not have significant overlap with known aging-related genes of other studies. Finally, the complement of DyNetAge has significant overlap with the complements of BrainExpression2004Age, BrainExpression2008Age and ADExpressionAge. Furthermore, the complement of DyNetAge has marginally significant overlap with the complement of HGPSExpressionAge. Though the complement of DyNetAge does not have significant overlap with the complement of SequenceAge, the complement of Brain08Expression also does not have significant overlap with the complement of SequenceAge. 

\textbf{Comment: Literature validation lacks a control. On p.7, the authors claim to validate an aging connection for up to 76\% of their predicted genes via automatic text mining of Pubmed. But for this to be meaningful, they have to report what percentage of non-aging-related genes would be similarly "validated" using the same search criteria.}

\textbf{Response:} We would like to thank the reviewer for pointing this. Previously we were able to validate 76\% genes of DyNetAge using automated literature mining meaning that 76\% genes of DyNetAge are found in at least one PubMed article that contains one of the aging-related keywords: `age', `aging' and `ageing'. Now, we find that 76\% genes of the complement of DyNetAge are also validated as aging-related using the same criteria. We apply the same validation criteria on ``ground truth'' aging-related datasets and their complements. We find that 72\%, 82\%, 73\%, and 98\% of the genes in ADExpressionAge, BrainExpression2004Age, BrainExpression2008Age, and SequenceAge are validated as aging-related, respectively, whereas 77\%, 76\%, 79\%, and 75\% of the genes in their complements are validated as aging-related. A higher percentage of genes of SequenceAge are validated as aging-related compared to other datasets because most of the genes of SequenceAge are annotated in PubMed articles. We observe a high percentage validation rate (75\%) in the complement of SequenceAge. Based on the observation, we argue that it is meaningless to perform automated literature validation to validate aging-related genes. Therefore, we will discard this analysis from the paper. 

\textbf{Comment: Do enrichment analysis lacks a control. On p.7, the authors report that 8 diseases are enriched in DyNetAge genes, and 14 diseases are enriched in non-aging-related genes. They make some tenuous connections between these diseases and aging, arguing that these findings support the aging evidence of DyNetAge genes. But what are the 14 diseases significant in the background set, and how many of these are similarly aging-related?}

\textbf{Response:} To answer this question, we identify diseases that are enriched in each aging-related dataset and its complement and compute the overlap of statistical significance between diseases that are enriched. We find that diseases enriched in DyNetAge are significantly overlapped with the diseases enriched in both BrainExpression2004Age (p-value of 0.03) and BrainExpression2008Age (p-value of 0.007). Therefore, DyNetAge is capable of capturing disease information enriched in brain aging-related expression data. On the other hand, diseases enriched in SequenceAge do not overlap with the diseases enriched in BrainExpression2008Age. Therefore, SequenceAge is not always capable of capturing disease information enriched in brain aging-related expression data.

Then, we compute the overlap of diseases that are enriched in DyNetAge with diseases that are enriched in the complement of ``ground truth'' datasets to see whether aging-related diseases revealed by DyNetAge are revealed as non-aging related by other datasets. A non-significant overlap is desired in this case. We find that diseases enriched in DyNetAge do not significantly overlap with diseases enriched in the complement of ``ground truth'' datasets. This outcome increases the credibility of the aging-related predictions in DyNetAge.

Then, we compute the overlap of diseases enriched in the complement of DyNetAge with diseases enriched in ``ground truth'' datasets to see whether non-aging-related diseases identified from DyNetAge are identified as aging-related by other datasets. A non-significant overlap is desired but a significant overlap is not necessarily discouraging. We find that diseases enriched in the complement of DyNetAge do not significantly overlap with diseases enriched in ``ground truth'' datasets only except SequenceAge. We argue that significant overlap between diseases enriched in the complement of DyNetAge and diseases enriched in SequenceAge is not entirely unexpected for some reasons. First, we do not claim that all genes in the complement of DyNetAge are non-aging-related. Rather we say that DyNetAge contains genes that we predict as aging-related and it is possible that some of the genes in the complement of DyNetAge could be identified as aging-related by other studies. This statement is also true for all ``ground truth'' aging-related datasets. Second, a large number of diseases (144, in particular) are enriched SequenceAge and it is expected that not all these diseases are truly aging-related. We find that 11 out of 14 diseases enriched in the complement of DyNetAge overlap with diseases enriched in SequenceAge. We perform manual literature validation on these 11 diseases and reveal that many of these diseases are not aging-related. For instance, asthma, which is enriched in SequenceAge, is common in all age groups \citep{Vignola2003Asthma}. Though Atherosclerosis is more frequent at old age, it is not an uncommon disease at young age \citep{Fausto1998Atherosclerosis}. Dermatitis is a skin disease that is common in all age groups \citep{Marks2003Dermatitis}. Obesity is not an aging-related disease \citep{Mendes2012Obesity}. Ulcerative colitis is most frequent in 15-30 years of age, but any age group can be affected by this disease \citep{Garud2009Ulcerative}. By discarding these diseases, we find that the overlap between the diseases enriched in the complement of DyNetAge and in SequenceAge is not statistically significant (p-value of 0.14).

\textbf{Comment: It's not clear whether any multiple testing correction is used when determining P values for each gene.}

\textbf{Response:} We argue that multiple test correction is not applicable in determining p-value for each gene. Unlike microarray study, we do not experimentally determine whether a gene is aging-related. Rather we predict aging-related genes and assign a prediction score indicating the credibility of the genes to be aging-related.  

\textbf{Comment: It's not clear how genes unexpressed in some samples are dealt with. Are the centrality values for these genes treated as NAs in samples where they are unexpressed?}

\textbf{Response:} The centrality value of an unexpressed gene is treated as zero.

\textbf{Comment: p.5, ``Relationships of different node centralities'': It's a bit surprising that 43\% of genes included in DyNetAge were significantly age-related for only one type of centrality. Is the subset of genes significant by multiple centrality measures more related to aging (i.e. is there a more significant overlap with the ``ground truth'' sets)? Are the genes that are only significant by a single centrality measure nearly significant (i.e., have low P values) by the other measures?}

\textbf{Response:} NEEDS TO EXPLAIN THE FIGURES

\begin{figure*}[htb]
\centering
\subfigure[]
{
	\includegraphics[scale=0.3]{figures/overlap-br04}
	\label{}
}
\subfigure[]
{
	\includegraphics[scale=0.3]{figures/overlap-br08}
	\label{}
}
\subfigure[]
{ 
	\includegraphics[scale=0.3]{figures/overlap-ad} 
	\label{} 
}
\subfigure[]
{
	\includegraphics[scale=0.3]{figures/overlap-hg}
	\label{}
}
\subfigure[]
{
	\includegraphics[scale=0.3]{figures/overlap-ga}
	\label{}
}
\vspace{-0.6cm}
\caption{The number of our predicted aging-related genes identified by each of seven node centralities individually (BETWC,
CLOSEC, CLUSC, DEGC, ECC, GDC and KC) or by at least one of them (UNION) and their overlap with ``ground truth'' aging-related datasets: (a) ADExpressionAge, (b) BrainExpression2004Age, (c) BrainExpression2008Age, (d) HGPSExpressionAge, and (e) SequenceAge.}
\label{fig:overlap-all}
\end{figure*}

\begin{figure*}[htb]
\centering
\subfigure[]
{ 
	\includegraphics[scale=0.3]{figures/noofcent-ad} 
	\label{} 
}
\subfigure[]
{
	\includegraphics[scale=0.3]{figures/noofcent-br04}
	\label{}
}
\subfigure[]
{
	\includegraphics[scale=0.3]{figures/noofcent-br08}
	\label{}
}
\subfigure[]
{
	\includegraphics[scale=0.3]{figures/noofcent-hg}
	\label{}
}
\subfigure[]
{
	\includegraphics[scale=0.3]{figures/noofcent-ga}
	\label{}
}
\vspace{-0.6cm}
\caption{The number of our predicted aging-related genes identified by exactly $k$ node centralities ($k=1,2,...,7$) and their overlap with ``ground truth'' aging-related datasets: (a) ADExpressionAge, (b) BrainExpression2004Age, (c) BrainExpression2008Age, (d) HGPSExpressionAge, and (e) SequenceAge.}
\label{fig:noofcent-all}
\end{figure*}

\begin{figure*}[htb]
\centering
\includegraphics[scale=0.3]{figures/pv-dist}
\vspace{-0.45cm}
\caption{Distribution of p-values of our predicted aging-related genes identified by exactly $k$ node centralities ($k=1,2,...,7$).}
\label{fig:pv-dist}
\end{figure*}

\textbf{Comment: p.7, ``GO enrichment'': How many GO terms were significant in total for SequenceAge and BrainExpression2004Age?}

\textbf{Response:} 762 and 251 GO terms are enriched in SequenceAge and BrainExpression2004Age, respectively.

\textbf{Comment: Each probe is scored in a binary way, expressed or not, using $p=.04$ as a threshold. It is not said precisely how that decision is make.}

\textbf{Response:} It is one of the standard ways in microarray study to pick p-value of 0.04 to determine whether a gene is expressed. The actual gene expression data, used in this study, has 

\textbf{Comment: The paper has some serious limitations. First only 30 individuals are used and they are from a nearly 10 year old data set. That is not a sufficient basis for supporting claims about aging in general. Each age-specific network is based on just one individual.}

\textbf{Response:} NEEDS TO WRITE THAT MICROARRAY EXPERIMENT TECHNOLOGY HAS NOT CHANGED DRASTICALLY DURING LAST TEN YEARS

\textbf{Comment: Second, space limitations and a non-mathematical presentation mean that the reader can not be sure what was computer. The wording is ambiguous in many places leaving readers to guess about what computation was done.}

\textbf{Response:} It is a writing issue.

\textbf{Comment: There should be some discussion of the biological implications of each statistic (e.g. centrality or clustering coefficient). It is not clear why ones in the paper were chosen.}

\textbf{Response:} It is a writing issue.

\fi

\paragraph{Funding\textcolon} NSF EAGER CCF-1243295 grant.

\vspace{-0.7cm}

\bibliographystyle{bioinformatics}
\bibliography{document}

\begin{thebibliography}{50}
\expandafter\ifx\csname natexlab\endcsname\relax\def\natexlab#1{#1}\fi
\expandafter\ifx\csname url\endcsname\relax
  \def\url#1{\texttt{#1}}\fi
\expandafter\ifx\csname urlprefix\endcsname\relax\def\urlprefix{URL }\fi

\bibitem[{Aragues \emph{et~al.}(2008)Aragues, Sander and Oliva}]{Aragues2008}
Aragues, R., Sander, C. and Oliva, B. (2008) Predicting cancer involvement of
  genes from heterogeneous data.
\newblock \emph{BMC Bioinformatics}, \textbf{9}, 172.

\bibitem[{Ashburner \emph{et~al.}(2000)Ashburner, Ball, Blake, Botstein,
  Butler, Cherry, Davis, Dolinski, Dwight, Eppig, Harris, Hill, Issel-Tarver,
  Kasarskis, Lewis, Matese, Richardson, Ringwald, Rubin and
  Sherlock}]{GENEONTOLOGY}
Ashburner, M. \emph{et~al.} (2000) Gene ontology: tool for the unification of
  biology.
\newblock \emph{Nature Genetics}, \textbf{25}, 25--29.

\bibitem[{Barab\'{a}si and Oltvai(2004)}]{Barabasi_Oltvai04}
Barab\'{a}si, A.~L. and Oltvai, Z. (2004) Network biology: Understanding the
  cell's functional organization.
\newblock \emph{Nature Reviews}, \textbf{5}, 101--113.

\bibitem[{Berchtold \emph{et~al.}(2008)Berchtold, Cribbs, Coleman, Rogers,
  Head, Kim, Beach, Miller, Troncoso, Trojanowski, Zielke and
  Cotman}]{Berchtold2008}
Berchtold, N.~C. \emph{et~al.} (2008) Gene expression changes in the course of
  normal brain aging are sexually dimorphic.
\newblock \emph{PNAS}, \textbf{105}, 15605--10.

\bibitem[{Breitkreutz \emph{et~al.}(2008)Breitkreutz, Stark, , Reguly, Boucher,
  Breitkreutz, Livstone, Oughtred, Lackner, Bahler, Wood, Dolinski and
  Tyers}]{BIOGRID}
Breitkreutz, B.~J. \emph{et~al.} (2008) {T}he {B}io{GRID} {I}nteraction
  {D}atabase: 2008 update.
\newblock \emph{Nucleic Acids Research}, \textbf{36}, D637--D640.

\bibitem[{de~Magalh\~{a}es(2009)}]{Magalhaes2009}
de~Magalh\~{a}es, J. (2009) Aging research in the post-genome era: New
  technologies for an old problem.
\newblock In Foyer, C., Faragher, R. and Thornalley, P. (eds.), \emph{Redox
  Metabolism and Longevity Relationships in Animals and Plants}, pp. 99--115.
  Taylor and Francis, New York.

\bibitem[{de~Magalh\~{a}es \emph{et~al.}(2009)de~Magalh\~{a}es, Budovsky,
  Lehmann, Costa, Li, Fraifeld and Church}]{Magalhaes2009a}
de~Magalh\~{a}es, J. \emph{et~al.} (2009) {The Human Ageing Genomic Resources:
  online databases and tools for biogerontologists.}
\newblock \emph{Aging Cell}, \textbf{8}, 65--72.

\bibitem[{Du \emph{et~al.}(2009)Du, Feng, Flatow, Song, Holko, Kibbe and
  Lin}]{Du2009}
Du, P. \emph{et~al.} (2009) From disease ontology to disease-ontology lite:
  statistical methods to adapt a general-purpose ontology for the test of
  gene-ontology associations.
\newblock \emph{Bioinformatics}, \textbf{25}, i63--68.

\bibitem[{Dyer \emph{et~al.}(2008)Dyer, Murali and Sobral}]{Dyer2008}
Dyer, M., Murali, T. and Sobral, B. (2008) The landscape of human proteins
  interacting with viruses and other pathogens.
\newblock \emph{PLoS Pathog}, \textbf{4}, e32+.

\bibitem[{Ferrarini \emph{et~al.}(2005)Ferrarini, Bertelli, Feala, McCulloch
  and Paternostro}]{aging1}
Ferrarini, L., Bertelli, L., Feala, J., McCulloch, A.~D. and Paternostro, G.
  (2005) A more efficient search strategy for aging genes based on
  connectivity.
\newblock \emph{Bioinformatics}, \textbf{21}, 338--348.

\bibitem[{Fortney \emph{et~al.}(2010)Fortney, Kotlyar and
  Jurisica}]{Fortney2010}
Fortney, K., Kotlyar, M. and Jurisica, I. (2010) {Inferring the functions of
  longevity genes with modular subnetwork biomarkers of Caenorhabditis elegans
  aging}.
\newblock \emph{Genome Biology}, \textbf{11}, R13+.

\bibitem[{Goh \emph{et~al.}(2007)Goh, Cusick, Valle, Childs, Vidal and
  Barab\'{a}si}]{Goh2007}
Goh, K. \emph{et~al.} (2007) {The human disease network}.
\newblock \emph{PNAS}, \textbf{104}, 8685--8690.

\bibitem[{Ho \emph{et~al.}(2010)Ho, Milenkovi\'{c}, Memisevic, Aruri, Przulj
  and Ganesan}]{Ho2010}
Ho, H. \emph{et~al.} (2010) Protein interaction network uncovers melanogenesis
  regulatory network components within functional genomics datasets.
\newblock \emph{BMC Systems Biology}, \textbf{4}.

\bibitem[{Janji\'{c} and Pr\v{z}ulj(2012)}]{Janjic2012}
Janji\'{c}, V. and Pr\v{z}ulj, N. (2012) The core diseasome.
\newblock \emph{Molecular bioSystems}, \textbf{8}, 2614--25.

\bibitem[{Jeong \emph{et~al.}(2001)Jeong, Mason, Barab\'{a}si and
  Oltvai}]{Jeong01}
Jeong, H., Mason, S.~P., Barab\'{a}si, A.~L. and Oltvai, Z.~N. (2001) Lethality
  and centrality in protein networks.
\newblock \emph{Nature}, \textbf{411}, 41--2.

\bibitem[{Jonsson and Bates(2006)}]{JB06}
Jonsson, P.~F. and Bates, P.~A. (2006) Lobal topological features of cancer
  proteins in the human interactome.
\newblock \emph{Bioinformatics}, \textbf{22}, 2291--2297.

\bibitem[{Kitsak \emph{et~al.}(2010)Kitsak, Gallos, Havlin, Liljeros, Muchnik,
  Stanley and Makse}]{Kitsak2010}
Kitsak, M. \emph{et~al.} (2010) Identification of influential spreaders in
  complex networks.
\newblock \emph{Nature Physics}, \textbf{6}, 888--893.

\bibitem[{Kosch\"{u}tzki and Schreiber(2008)}]{Koschutzki2008}
Kosch\"{u}tzki, D. and Schreiber, F. (2008) Centrality analysis methods for
  biological networks and their application to gene regulatory networks.
\newblock \emph{Gene Regulation and Systems Biology}, \textbf{2}, 193--201.

\bibitem[{Kriete \emph{et~al.}(2011)Kriete, Lechner, Clearfield and
  Bohmann}]{Kriete2011}
Kriete, A., Lechner, M., Clearfield, D. and Bohmann, D. (2011) {Computational
  systems biology of aging}.
\newblock \emph{Wiley Interdiscip Rev Syst Biol Med}, \textbf{3}, 414--28.

\bibitem[{Kuchaiev \emph{et~al.}(2011)Kuchaiev, Stevanovi\'{c}, Hayes and
  Pr\v{z}ulj}]{GraphCrunch2}
Kuchaiev, O., Stevanovi\'{c}, A., Hayes, W. and Pr\v{z}ulj, N. (2011)
  Graph{C}runch 2: Software tool for network modeling, alignment and
  clustering.
\newblock \emph{BMC Bioinformatics}, \textbf{12}.

\bibitem[{Liu \emph{et~al.}(2011)Liu, Barkho, Ruiz, Diep, Qu, Yang, Panopoulos,
  Suzuki, Kurian, Walsh, Thompson, Boue, Fung, Sancho-Martinez, Zhang, Yates,
  Carlos and Belmonte}]{Liu2011}
Liu, G.-H. \emph{et~al.} (2011) Recapitulation of premature ageing with ipscs
  from hutchinson–gilford progeria syndrome.
\newblock \emph{Nature}, \textbf{472}, 221--5.

\bibitem[{Lu \emph{et~al.}(2004)Lu, Pan, Kao, Li, Kohane, Chan and
  Yankner}]{Lu2004}
Lu, T. \emph{et~al.} (2004) {Gene regulation and DNA damage in the ageing human
  brain}.
\newblock \emph{Nature}, \textbf{429}, 883--891.

\bibitem[{Memisevi\'{c} \emph{et~al.}(2010)Memisevi\'{c}, Milenkovi\'{c} and
  Pr\v{z}ulj}]{Memisevic10a}
Memisevi\'{c}, V., Milenkovi\'{c}, T. and Pr\v{z}ulj, N. (2010) An integrative
  approach to modeling biological networks.
\newblock \emph{Journal of Integrative Bioinformatics}, \textbf{7}, 120.

\bibitem[{Memi\v{s}evi\'{c} \emph{et~al.}(2010)Memi\v{s}evi\'{c},
  Milenkovi\'{c} and Pr\v{z}ulj}]{Memisevic10b}
Memi\v{s}evi\'{c}, V., Milenkovi\'{c}, T. and Pr\v{z}ulj, N. (2010)
  Complementarity of network and sequence information in homologous proteins.
\newblock \emph{Journal of Integrative Bioinformatics}, \textbf{7}, 135.

\bibitem[{Milenkovi\'{c} \emph{et~al.}(2009)Milenkovi\'{c}, Filippis, Lappe and
  Pr\v{z}ulj}]{Milenkovic2009}
Milenkovi\'{c}, T., Filippis, I., Lappe, M. and Pr\v{z}ulj, N. (2009) Optimized
  null model for protein structure networks.
\newblock \emph{PLoS ONE}, \textbf{4}, e5967.

\bibitem[{Milenkovi\'{c} \emph{et~al.}(2008)Milenkovi\'{c}, Lai and
  Pr\v{z}ulj}]{GraphCrunch}
Milenkovi\'{c}, T., Lai, J. and Pr\v{z}ulj, N. (2008) Graph{C}runch: a tool for
  large network analyses.
\newblock \emph{BMC Bioinformatics}, \textbf{9}.

\bibitem[{Milenkovi\'{c} \emph{et~al.}(2010{\natexlab{a}})Milenkovi\'{c},
  Memisevi\'{c}, Ganesan and Pr\v{z}ulj}]{MMGP_Roy_Soc_09}
Milenkovi\'{c}, T., Memisevi\'{c}, V., Ganesan, A.~K. and Pr\v{z}ulj, N.
  (2010{\natexlab{a}}) Systems-level cancer gene identification from protein
  interaction network topology applied to melanogenesis-related interaction
  networks.
\newblock \emph{Journal of the Royal Society Interface}, \textbf{7}, 423--437.

\bibitem[{Milenkovi\'{c} \emph{et~al.}(2011)Milenkovi\'{c}, Memi\v{s}evi\'{c},
  Bonato and Pr\v{z}ulj}]{Milenkovic2011}
Milenkovi\'{c}, T., Memi\v{s}evi\'{c}, V., Bonato, A. and Pr\v{z}ulj, N. (2011)
  Dominating biological networks.
\newblock \emph{PLoS ONE}, \textbf{6}, e23016.

\bibitem[{Milenkovi\'{c} \emph{et~al.}(2010{\natexlab{b}})Milenkovi\'{c}, Ng,
  Hayes and Pr\v{z}ulj}]{HGRAAL}
Milenkovi\'{c}, T., Ng, W., Hayes, W. and Pr\v{z}ulj, N. (2010{\natexlab{b}})
  Optimal network alignment with graphlet degree vectors.
\newblock \emph{Cancer Informatics}, \textbf{9}, 121--137.

\bibitem[{Milenkovi\'{c} and Pr\v{z}ulj(2008)}]{Milenkovic2008}
Milenkovi\'{c}, T. and Pr\v{z}ulj, N. (2008) Uncovering biological network
  function via graphlet degree signatures.
\newblock \emph{Cancer Informatics}, \textbf{6}, 257--273.

\bibitem[{Motulsky(1995)}]{Motulsky1995}
Motulsky, H. (1995) \emph{Intuitive Biostatistics}.
\newblock Oxford University Press, 1 edition.

\bibitem[{Pastor-Satorras \emph{et~al.}(2003)Pastor-Satorras, Smith and
  Sole}]{PastorSatorras03}
Pastor-Satorras, R., Smith, E. and Sole, R.~V. (2003) Evolving protein
  interaction networks through gene duplication.
\newblock \emph{Journal of Theoretical Biology}, \textbf{222}, 199--210.

\bibitem[{Peri \emph{et~al.}(2004)Peri, Navarro, Kristiansen, Amanchy,
  Surendranath, Muthusamy, Gandhi, Chandrika, Deshpande, Suresh, Rashmi,
  Shanker, Padma, Niranjan, Harsha, Talreja, Vrushabendra, Ramya, Yatish, Joy,
  Shivashankar, Kavitha, Menezes, Choudhury, Ghosh, Saravana, Chandran, Mohan,
  Jonnalagadda, Prasad, Kumar-Sinha, Deshpande and Pandey}]{HPRD}
Peri, S. \emph{et~al.} (2004) Human protein reference database as a discovery
  resource for proteomics.
\newblock \emph{Nucleic Acids Res}, \textbf{32 Database issue}, D497--501.
\newblock 1362-4962Journal Article.

\bibitem[{Promislow(2004)}]{aging2}
Promislow, D. (2004) Protein networks, pleiotropy and the evolution of
  senescence.
\newblock \emph{Proceedings of the Royal Society B: Biological Sciences},
  \textbf{1545}, 1225--1234.

\bibitem[{Pr\v{z}ulj(2007)}]{Przulj06ECCB}
Pr\v{z}ulj, N. (2007) Biological network comparison using graphlet degree
  distribution.
\newblock \emph{Bioinformatics}, \textbf{23}, e177--e183.

\bibitem[{Pr\v{z}ulj(2011)}]{Przulj2011}
Pr\v{z}ulj, N. (2011) {Protein-protein interactions: Making sense of networks
  via graph-theoretic modeling}.
\newblock \emph{Bioessays}, \textbf{33}, 115--123.

\bibitem[{Pr\v{z}ulj \emph{et~al.}(2010)Pr\v{z}ulj, Kuchaiev, Stevanovi\'{c}
  and Hayes}]{GeoGD}
Pr\v{z}ulj, N., Kuchaiev, O., Stevanovi\'{c}, A. and Hayes, W. (2010) Geometric
  evolutionary dynamics of protein interaction networks.
\newblock \emph{Pacific Symposium on Biocomputing}, pp. 178--89.

\bibitem[{Przytycka and Kim(2010)}]{Przytycka2010}
Przytycka, T. and Kim, Y. (2010) {Network integration meets network dynamics}.
\newblock \emph{BMC Biology}, \textbf{8}, 48+.

\bibitem[{Radivojac \emph{et~al.}(2008)Radivojac, Peng, Clark, Peters, Mohan,
  Boyle and D.}]{Radivojac2008}
Radivojac, P. \emph{et~al.} (2008) An integrated approach to inferring
  gene-disease associations in humans.
\newblock \emph{Proteins}, \textbf{72}, 1030--1037.

\bibitem[{Ratmann \emph{et~al.}(2009)Ratmann, Wiuf and Pinney}]{Ratman2009}
Ratmann, O., Wiuf, C. and Pinney, J.~W. (2009) From evidence to inference:
  probing the evolution of protein interaction networks.
\newblock \emph{HFSP Journal}, \textbf{3}, 290--306.

\bibitem[{Reja \emph{et~al.}(2009)Reja, Venkatakrishnan, Lee, Kim, Ryu, Gong,
  Bhak and Park}]{Reja2009}
Reja, R. \emph{et~al.} (2009) {MitoInteractome: mitochondrial protein
  interactome database, and its application in 'aging network' analysis.}
\newblock \emph{BMC genomics}, \textbf{10 Suppl 3}, S20+.

\bibitem[{Scardoni \emph{et~al.}(2009)Scardoni, Petterlini and
  Laudanna}]{Scardoni2009}
Scardoni, G., Petterlini, M. and Laudanna, C. (2009) Analyzing biological
  network parameters with centiscape.
\newblock \emph{Bioinformatics}, \textbf{25}, 2857--2859.

\bibitem[{Sharan and Ideker(2008)}]{Sharan2008}
Sharan, R. and Ideker, T. (2008) Protein networks in disease.
\newblock \emph{Genome Research}, \textbf{18}, 644--652.

\bibitem[{Sharan \emph{et~al.}(2007)Sharan, Ulitsky and Shamir}]{Sharan2007}
Sharan, R., Ulitsky, I. and Shamir, R. (2007) Network-based prediction of
  protein function.
\newblock \emph{Molecular Systems Biology}, \textbf{3}, 1--13.

\bibitem[{S\H{o}ti and Csermely(2007)}]{Soti2007}
S\H{o}ti, C. and Csermely, P. (2007) Aging cellular networks: Chaperones as
  major participants.
\newblock \emph{Experimental Gerontology}, \textbf{42}, 113--119.

\bibitem[{Simpson \emph{et~al.}(2011)Simpson, Ince, Shaw, Heath, Raman,
  Garwood, Gelsthorpe, Baxter, Forster, Matthews, Brayne and
  Wharton}]{Simpson2011}
Simpson, J.~E. \emph{et~al.} (2011) Microarray analysis of the astrocyte
  transcriptome in the aging brain: relationship to alzheimer's pathology and
  apoe genotype.
\newblock \emph{Neurobiology of Aging}, \textbf{32}, 1795--807.

\bibitem[{Soltow \emph{et~al.}(2010)Soltow, Jones and Promislow}]{Soltow2010}
Soltow, Q., Jones, D. and Promislow, D. (2010) {A Network Perspective on
  Metabolism and Aging.}
\newblock \emph{Integrative and comparative biology}, \textbf{50}, 844--854.

\bibitem[{Vanunu \emph{et~al.}(2010)Vanunu, Magger, Ruppin, Shlomi and
  Sharan}]{Sharan10}
Vanunu, O., Magger, O., Ruppin, E., Shlomi, T. and Sharan, R. (2010)
  Associating genes and protein complexes with disease via network propagation.
\newblock \emph{PLoS Computational Biology}, \textbf{6}, e1000641.

\bibitem[{Wieser \emph{et~al.}(2011)Wieser, Papatheodorou, Ziehm and
  Thornton}]{Wieser2011}
Wieser, D., Papatheodorou, I., Ziehm, M. and Thornton, J. (2011) {Computational
  biology for ageing}.
\newblock \emph{Philosophical Transactions of the Royal Society B: Biological
  Sciences}, \textbf{366}, 51--63.

\bibitem[{Wuchty and Stadler(2003)}]{Wuchty2003}
Wuchty, S. and Stadler, P.~F. (2003) Centers of complex networks.
\newblock \emph{Journal of Theoretical Biology}, \textbf{223}, 45--53.

\end{thebibliography}

\end{document}